\documentclass[journal,draftclsnofoot,onecolumn,12pt]{IEEEtran}

\usepackage{amsthm,amssymb,graphicx,multirow,amsmath,color,amsfonts}
\usepackage[update,prepend]{epstopdf}
\usepackage[noadjust]{cite}
\usepackage[latin1]{inputenc}
\usepackage{tikz}
\usetikzlibrary{arrows,calc} 
\usepackage{bbm} 
\usepackage{pdfpages}
\usepackage{tabulary}
\usepackage{multirow}
\usepackage{comment}

\def\nbo{{\mathbf{o}}}
\def\nbp{{\mathbf{p}}}

\def\nbx{{\mathbf{x}}}
\def\nby{{\mathbf{y}}}

\def\nb0{{\mathbf{0}}}
\def\nb1{{\mathbf{1}}}

\def\ncalA{{\mathcal{A}}}
\def\ncalB{{\mathcal{B}}}

\def\ncalI{{\mathcal{I}}}
\def\ncalJ{{\mathcal{J}}}

\def\ncalN{{\mathcal{N}}}

\def\ncalR{{\mathcal{R}}}

\def\ncalX{{\mathcal{X}}}

\def\nbbE{{\mathbb{E}}}

\def\nbbP{{\mathbb{P}}}

\def\nbbR{{\mathbb{R}}}

\newtheorem{lemma}{Lemma}

\newtheorem{definition}{Definition}

\newtheorem{theorem}{Theorem}
\newtheorem{prop}{Proposition}
\newtheorem{cor}{Corollary}

\newtheorem{remark}{Remark}

\allowdisplaybreaks 

\begin{document}
\title{Performance Characterization of Canonical Mobility Models in Drone Cellular Networks}
\author{
Morteza Banagar and Harpreet S. Dhillon
\thanks{The authors are with Wireless@VT, Department of ECE, Virginia Tech, Blacksburg, VA. Email: \{mbanagar, hdhillon\}@vt.edu. The support of the US NSF (Grant CNS-1617896) is gratefully acknowledged. This paper will be presented in part at the 2019 IEEE Globecom in Waikoloa, HI, USA \cite{C_Morteza_3GPP_2019, C_Morteza_Fundamentals_2019}. \hfill Manuscript updated: \today.}}

\maketitle

\vspace{-1.7cm}
\begin{abstract}
In this paper, we characterize the performance of several canonical mobility models in a drone cellular network in which drone base stations (DBSs) serve user equipments (UEs) on the ground. In particular, we consider the following four mobility models: (i) straight line (SL), (ii) random stop (RS), (iii) random walk (RW), and (iv) random waypoint (RWP), among which the SL mobility model is inspired by the simulation models used by the third generation partnership project (3GPP) for the placement and trajectory of drones, while the other three are well-known canonical models (or their variants) that offer a useful balance between realism and tractability. Assuming the nearest-neighbor association policy, we consider two service models for the UEs: (i) {\em UE independent model} (UIM), and (ii) {\em UE dependent model} (UDM). While the serving DBS follows the same mobility model as the other DBSs in the UIM, it is assumed to fly towards the UE of interest in the UDM and hover above its location after reaching there. The main contribution of this paper is a unified approach to characterize the point process of DBSs for all the mobility and service models. Using this, we provide exact mathematical expressions for the average received rate and the session rate as seen by the typical UE. Further, using tools from calculus of variations, we concretely demonstrate that the simple SL mobility model provides a lower bound on the performance of other general mobility models (including the ones in which drones follow {\em curved} trajectories) as long as the movement of each drone in these models is independent and identically distributed (i.i.d.). To the best of our knowledge, this is the first work that provides a rigorous analysis of key canonical mobility models for an infinite drone cellular network and establishes useful connections between them.
\end{abstract}


\vspace{-0.4cm}
\section{Introduction} \label{sec:intro}
Wireless networks are all set to undergo a major transformation from being predominantly terrestrial to the ones that will have an elaborate and dynamic aerial component in the form of drone networks \cite{J_Zeng_Wireless_2016}. Owing to their deployment flexibility, drones are currently being considered for a variety of use cases, such as acting as mobile relays to expand the coverage of cellular networks, providing network connectivity for public safety applications, and setting up temporary networks in the times of natural disasters or large social gatherings. While the flexibility offered by the mobility of drones is highly appealing to the network designers, it also adds an entirely new dimension to the system design that was not present in the traditional terrestrial networks. In particular, while mobility is known to have a fundamental impact on the system-level performance of wireless networks, e.g., see \cite{J_Grossglauser_Mobility_2002}, the research focus in this direction has traditionally been on the setting in which the UEs are mobile while the base stations (BSs) are static \cite{J_Bettstetter_Node_2003, J_Lin_Towards_2013}. This is clearly not the case in drone-assisted communication networks where some mobile drones may act as BSs \cite{J_Mozaffari_Tutorial_2018}. Not surprisingly, the support of mobile drones, either as UEs or BSs, has also been recently explored in 3GPP studies \cite{3gpp_36777, 3gpp_22829}. In general, there has been an increasing interest in the community to characterize the effect of drone mobility on the design and performance of drone-assisted cellular networks. Due to the irregularity of the drone placements and their trajectories, it is expected that powerful tools from stochastic geometry and point process theory could be leveraged for accurate modeling and tractable analysis of drone networks. Inspired by this, the main focus of this paper is to present a unified performance analysis of a DBS network under various mobility models that involves a novel characterization of the underlying point processes as a function of time.

\vspace{-0.4cm}
\subsection{Related Works}  \label{sec:related}
This paper builds on the following two key lines of research: (i) stochastic geometry for drone networks, and (ii) mobility models in wireless networks. Although sparse, there are some works that lie at their intersection and will be discussed below after establishing the prior art for each research direction separately.

\emph{Stochastic Geometry for Drone Networks.}
Owing to its ability to capture irregularity in the placement and movement of drones, stochastic geometry has recently found many applications in the performance analysis of drone networks \cite{J_Chetlur_Downlink_2017, C_Chetlur_Downlink_2016, J_Zhang_Spectrum_2017, J_Wang_Modeling_2018, C_Alzenad_Coverage_2018, J_Mozaffari_Unmanned_2016, J_Hayajneh_Performance_2018}. In \cite{J_Chetlur_Downlink_2017}, the authors considered a finite network of DBSs distributed as a uniform binomial point process (BPP) \cite{J_Afshang_Fundamentals_2017} and derived the coverage probability of the network. In \cite{J_Zhang_Spectrum_2017}, the problem of spectrum sharing for a network of drone small cells as an underlay to a conventional cellular network has been studied. On the similar lines, the authors of \cite{J_Wang_Modeling_2018} investigated the coexistence of BSs and DBSs using probabilistic line-of-sight (LoS) and non-line-of-sight (NLoS) propagation models \cite{J_AlHourani_Optimal_2014}, where the locations of BSs and DBSs are modeled as a superposition of a Poisson point process (PPP) and a BPP. The work presented in \cite{C_Alzenad_Coverage_2018} considered a network of DBSs modeled as a PPP serving ground UEs. In particular, incorporating LoS and NLoS propagation models, the authors derived approximations for the coverage probability and the received rate in the network. In \cite{J_Mozaffari_Unmanned_2016}, the coexistence between a single DBS and an underlaid device-to-device (D2D) network has been analyzed in terms of coverage probability and rate. Motivated by the requirement to provide coverage in a post-disaster scenario, the authors in \cite{J_Hayajneh_Performance_2018} analyzed the performance of a DBS network serving clustered UEs.

\emph{Mobility Models in Wireless Networks.}
Mobility modeling is a well-established area of research in wireless networks \cite{J_Camp_Survey_2002, C_Bai_Survey_2004, J_Gong_Interference_2014, J_Tabassum_Fundamentals_2019}. Perhaps the simplest mobility model is the one in which nodes move along straight lines in random directions with a constant speed. Variants of this simple model have been used extensively in the literature, e.g., see \cite{J_Neely_Capacity_2005, C_Kong_Latency_2008, J_Gong_Interference_2014, J_Madadi_Shared_2018, 3gpp_36777}. Although this model may appear simplistic, it is known to provide performance bounds and useful insights in wireless networks \cite{J_Neely_Capacity_2005}. In fact, this model has also been used recently to model drone mobility in 3GPP studies related to drone networks \cite{3gpp_36777}. Among other mobility models that have been studied in wireless networks (such as RW, RWP, random direction, Brownian motion, Levy walk, and Gauss-Markov), RW and RWP mobility models have been more popular because of their tractability \cite{C_McGuire_Stationary_2005, J_Bettstetter_Node_2003, J_Groenevelt_Relaying_2006, J_Rhee_Levy_2011, J_Camp_Survey_2002}. Hence, we also study these two models in this paper.

In the RW model, the mobile nodes can change their directions \cite{J_Camp_Survey_2002, C_McGuire_Stationary_2005, C_Kong_Latency_2008}. In particular, each node selects a uniformly random direction and a random speed in each time slot and moves along a straight line using the selected direction and speed. Upon arrival at the destination, it repeats this procedure. As a generalization of the RW model, the finite RWP model was proposed in \cite{B_Johnson_Dynamic_1996} by adding a random pause time between direction and/or speed changes in the RW model. Specifically, the mobility process in the finite RWP model starts with a random pause time at the initial waypoint of a node in a finite region. Then, a destination waypoint is selected randomly in that region and the node moves to this destination waypoint along a straight line with a random speed. Upon its arrival, the node pauses for another random time and repeats this procedure. Due to some major drawbacks of this model, such as nonuniform distribution of the nodes \cite{J_Bettstetter_Node_2003, J_Bettstetter_Stochastic_2004, J_Hyytia_Spatial_2006}, the authors in \cite{J_Lin_Towards_2013} extended the finite RWP to an infinite RWP model, where the nodes are allowed to move over the entire plane. In particular, each node first pauses for a random time at its initial waypoint and determines the next waypoint by choosing a uniformly random direction and a random transition distance. It then moves towards the chosen waypoint at a random speed, pauses at this new waypoint for another random time, and repeats this procedure.

\emph{Mobile Drone Networks.}
Among many works on the analysis of drone networks, only a handful of them considered \emph{mobile} drones. Considering a mobile network of DBSs that are initially modeled as a BPP, the authors in \cite{J_Enayati_mobile_2018} used results from \cite{J_Chetlur_Downlink_2017} to design stochastic trajectory processes for the mobility of DBSs that provide the same coverage as given in \cite{J_Chetlur_Downlink_2017}. In \cite{J_Sharma_Coverage_2019} and \cite{J_Sharma_Random_2019}, the authors modeled the motion of DBSs in a 3D finite network by the finite RWP and RW models, and using the results of \cite{J_Bettstetter_Node_2003}, derived the coverage probability of the network.

\vspace{-0.2cm}
\subsection{Contributions} \label{sec:contributions}
This paper develops a unified approach for the performance analysis of drone cellular networks under several key canonical models. In particular, we model the initial locations of a mobile network of DBSs as a homogeneous PPP operating at a constant height that serve UEs on the ground. The serving DBS is selected based on the nearest-neighbor association policy and all the other DBSs are regarded as interferers. We consider four canonical mobility models for the interfering DBSs, i.e., SL, RS, RW, and RWP, and two service models for the serving DBS: (i) UIM, in which the serving DBS follows the same mobility model as the interfering DBSs, and (ii) UDM, in which the serving DBS moves towards the typical UE at a constant height and keeps hovering above its location after reaching there until its transmission is completed. For this setup, our key contributions are described next.

\subsubsection{Distributional Properties of Canonical Mobility Models}
We derive various distributional properties for all the mobility models. Most notably, for the RW and RWP, we derive the joint distribution of the net displacement of a DBS and its total traveled distance at each waypoint. Using this joint distribution, we compute the distribution of the net displacement of a DBS at any time $t$. We also provide insightful asymptotic results for these distributions.

\subsubsection{Unified Framework for Characterizing the Point Process of DBSs}
We present a novel characterization of the point process of DBSs as seen by the typical UE at any time $t$ for all the service and mobility models. Since the displacement of each DBS is independent from the others, we apply displacement theorem from stochastic geometry along with the net displacement results described above to characterize these point processes. Using this, we characterize the aggregate interference and key performance metrics, such as average and session rates, as seen by the typical UE for all the service and mobility models. 

\subsubsection{Establishing Connections among Mobility Models}
Borrowing tools from the calculus of variations, we establish meaningful connections among our mobility models in terms of network performance. Specifically, we prove that the SL mobility model provides a lower bound on the average received rate over the space of all i.i.d. mobility models in which the drones can also follow curved trajectories (as long as they are i.i.d.). 

\vspace{-0.1cm}
\section{System Model} \label{sec:SysMod}
\vspace{-0.4cm}
\subsection{Spatial Setup} \label{subsec1:SpatialSetup}
\begin{figure}
\centering
\includegraphics[width=0.7\columnwidth]{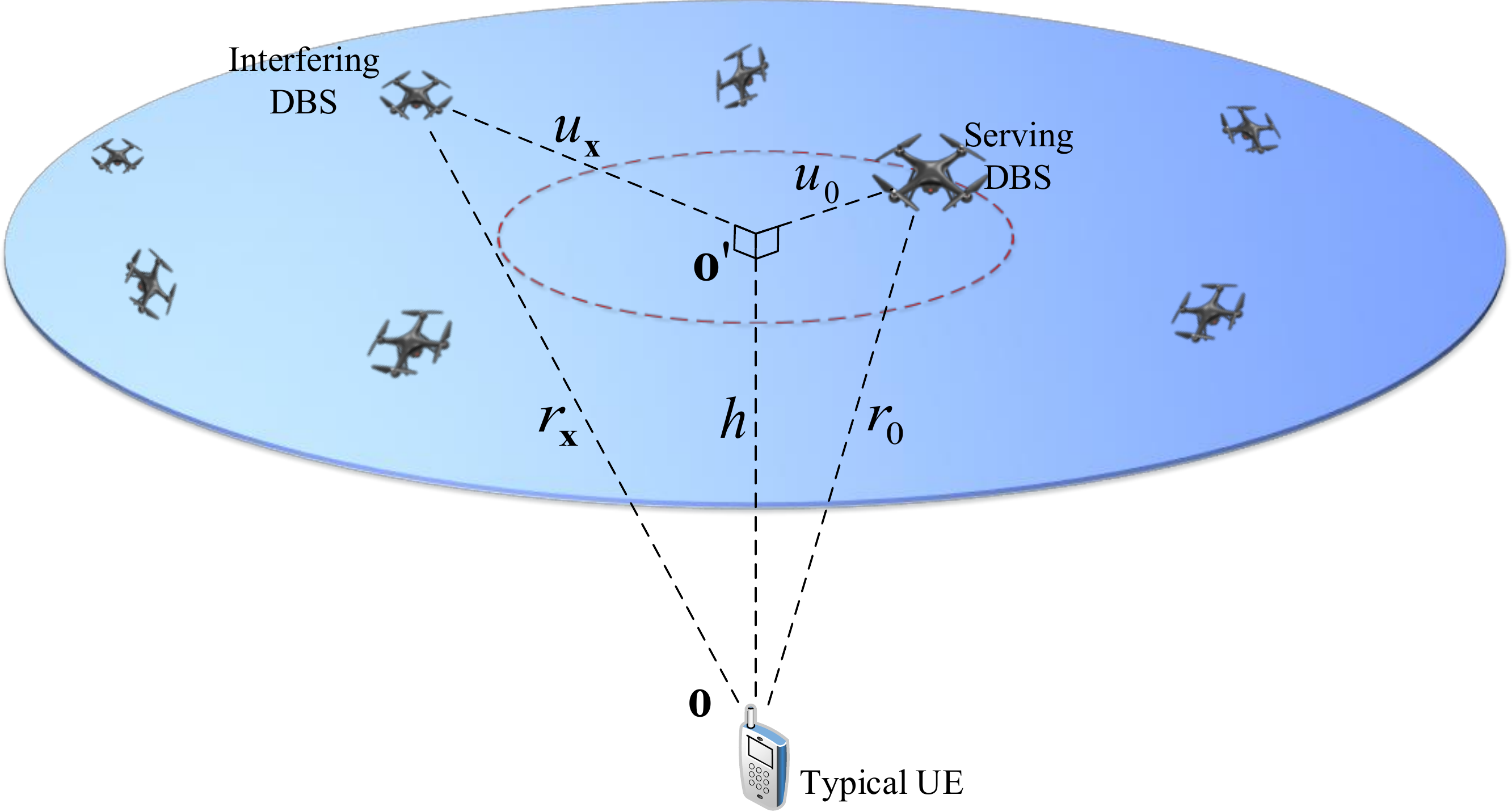}
\vspace{-0.6cm}
\caption{An illustration of the system model.}
\vspace{-0.6cm}
\label{fig:SystemModel_1}
\end{figure}

We consider a network of mobile DBSs serving UEs on the ground. We assume that the DBSs are located at a constant height $h$ from the ground and the temporal evolution of the DBS locations is modeled as the sequence of point processes $\Phi_{\rm D}(t) \subset \nbbR^2$, indexed by $t\in \nbbR^+$. Further, we assume that the initial locations of the DBSs are distributed as a homogeneous PPP with density $\lambda_0$, i.e., $\Phi_{\rm D}(0) \sim {\rm PPP}(\lambda_0)$. Terrestrial UEs are distributed as an independent homogeneous PPP $\Phi_{\rm U}$ on the ground. We also assume that the origin $\nbo = (0,0,0)$ of the 3D coordinate system is located on the ground, which is assumed to be aligned with the $xy$-plane. Throughout the paper, we refer to the $z=h$ plane as the DBS plane. In this setup, the projection of $\nbo$ onto the DBS plane is denoted by $\nbo' = (0, 0, h)$. Without loss of generality, the analysis will be performed for the {\em typical UE} placed at $\nbo$. As shown in Fig. \ref{fig:SystemModel_1}, the distances of a DBS at time $t$ located at $\nbx(t)\in\Phi_{\rm D}(t)$ from $\bf{o'}$ and $\bf{o}$ are denoted by $u_\nbx(t) = \|\nbx(t) - \bf{o'}\|$ and $r_\nbx(t) = \|\nbx(t) - {\bf o}\| = \sqrt{u_\nbx(t)^2+h^2}$, respectively. Moreover, the location of the nearest DBS to $\bf{o'}$ and its corresponding distance at time $t$ are denoted by $\nbx_0(t)$ and $u_0(t)$, respectively. Thus, the distance of the closest DBS to $\bf{o}$ at time $t$ is $r_0(t) = \sqrt{u_0(t)^2+h^2}$. For simplicity, we drop the time index $t$ for the distances defined at $t=0$, i.e., $u_0 \triangleq u_0(0)$, $r_0 \triangleq r_0(0)$, $u_\nbx \triangleq u_\nbx(0)$, and $r_\nbx \triangleq r_\nbx(0)$.

\vspace{-0.4cm}
\subsection{Service and Mobility Models} \label{subsec2:MobilityModels}
In this paper, we assume that each UE connects to its nearest DBS. For the typical UE, this DBS is called the \emph{serving DBS} while all the other DBSs are regarded as \emph{interfering DBSs}. In order to motivate the drone mobility models studied in this paper, we first take cue from the mobility model used in the 3GPP studies related to drone networks \cite{3gpp_36777}. In this model, the drones start their movement at randomly selected locations in the network and then move at a constant speed and height in straight lines and in uniformly random directions independently from each other for the entire duration of the simulation. It turns out that this simple enough model is already considered sufficient to capture key effects of drone mobility on the system-level performance. One can of course generalize this straight-line mobility model to arrive at classical canonical models, such as RW and RWP, which offer a useful balance between realism and tractability. Since all these models are important in their own right, we will develop a unified approach to analyze their performance jointly. Before that, we formally define the four mobility models considered in this paper next.
\begin{definition} \label{def:MobilityModels}
\emph {(Mobility Models).} We define the mobility models used in this paper as follows. Without loss of generality, we choose the $x$-axis as the reference for measuring all angles.
\begin{enumerate}
\item {\bf SL}: DBSs move constantly in random directions $\Theta \sim U[0, 2\pi)$ on straight lines with a constant speed $v$, independently from each other.
\item {\bf RS}: Each DBS moves a random distance $R \sim f_R(.)$ in a random direction $\Theta \sim U[0, 2\pi)$ on a straight line with a constant speed $v$, independently from the other DBSs, and then stops and hovers over the stopping location.
\item {\bf RW}: Each DBS selects a random direction $\Theta \!\sim\! U[0, 2\pi)$, independently of the other DBSs, and moves a random distance $R \sim f_R(.)$ in this direction with a constant speed $v$. Upon its arrival, it selects another random direction $\Theta$ and distance $R$ and repeats this procedure.
\item {\bf RWP}: In the beginning, each DBS hovers for a random time $T \!\sim\! f_T(.)$ at its initial location. It then selects a random direction $\Theta \!\sim\! U[0, 2\pi)$, independently of the other DBSs, and moves a random distance $R \sim f_R(.)$ in this direction with a constant speed $v$. Upon its arrival, it hovers for another random time $T$, sampled independently from $f_T(.)$, selects another random direction $\Theta$ and distance $R$, and repeats this procedure.
\end{enumerate}
\end{definition}
\begin{remark}
All the mobility models considered in this paper lie in the space of i.i.d. mobility models, where the drone trajectories are chosen independently of each other from the same common distribution. As will be clear in the sequel, some of our initial results hold for all i.i.d. models (including the ones with curved trajectories).
\end{remark}
\vspace{-0.3cm}
\begin{definition} \label{def:Flight}
\emph {(Flight).} In the RW and RWP mobility models, a flight is defined as the distance traveled by a DBS between two consecutive stop points.
\end{definition}
\vspace{-0.2cm}
For the serving DBS, we consider the following two service models.
\begin{enumerate}
\item UIM: The serving DBS follows the same mobility model as the interfering DBSs, independent of the typical UE location.
\item UDM: The serving DBS moves towards $\bf{o}'$ in the DBS plane and keeps hovering at this location until its transmission to the typical UE is completed.
\end{enumerate}
Note that in all the aforementioned models, the speeds of the serving and interfering DBSs are assumed to be the same. Moreover, since we are not considering any dependency across the user locations, DBS trajectories will be independent of each other in both the UIM and the UDM.
\vspace{-0.2cm}
\begin{remark}
In the UIM, since all DBSs move in different directions based on the defined mobility models, handover may occur. On the other hand, in the UDM, no matter what the mobility model is for the interfering DBSs, as long as the interfering DBSs move with the same speed as the serving DBS, the serving DBS will remain the closest DBS to the typical UE. Hence, based on our association policy, the serving DBS will not change in our mobility models, and thus, handover will not occur. Furthermore, UDM can be considered as a best-case model from the perspective of minimizing the distance between the typical UE and its serving DBS.
\end{remark}

\vspace{-0.4cm}
\subsection{Channel Model} \label{subsec3:ChannelModel}
We assume that all DBSs transmit with the same power level $P$ at all times. The received power at the typical UE from the serving DBS is assumed to be $P h_0(t) r_0(t)^{-\alpha}$, where $h_0(t)$ represents the small-scale fading power gain between the typical UE and the serving DBS and $\alpha>2$ is the path-loss exponent. Likewise, the interference power is $I(t)=\sum_{\nbx(t) \in \Phi_{\rm D}'(t)}P h_\nbx(t) r_\nbx(t)^{-\alpha}$, where $\Phi_{\rm D}'(t) \equiv \Phi_{\rm D}(t)\backslash \nbx_0(t)$ represents the point process of interfering DBSs and $h_\nbx(t)$ is the small-scale fading power gain between the typical UE and the interfering DBSs. Since the air-to-ground links may experience various fading scenarios, Nakagami-$m$ fading is used here to capture a large class of fading environments. The Nakagami-$m$ fading parameter is assumed to be $m_0$ and $m_\mathbf{x}$ for the serving and interfering links, respectively. Thus, the channel fading power gains $h_0(t)$ and $h_\mathbf{x}(t)$ follow gamma distributions with probability density function (pdf) $f_H(h) = \frac{m ^ m}{\Gamma(m)} h ^ {m - 1} {\rm e}^{-m h}$, where $\Gamma(x) = \int_0^\infty t^{x-1} {\rm e}^{-t} \,{\rm d}t$ is the gamma function. For the serving and interfering links, we assume $m = m_0$ and $m = m_\mathbf{x}$, respectively, and we consider integer values for $m_0$ and $m_\mathbf{x}$ for mathematical tractability.

\vspace{-0.4cm}
\subsection{Metrics of Interest} \label{subsec4:Metrics}
The received signal-to-interference ratio (${\rm SIR}$) at the typical UE is defined as
\begin{equation}
{\rm SIR}(t) = \frac{P h_0(t) r_0(t)^{-\alpha}}{I(t)}.
\end{equation}
We now define our ${\rm SIR}$-based performance metrics as follows.

\noindent\emph{Average rate}: Average received rate is given as $R(t) = \nbbE[\log\left(1 + {\rm SIR}(t)\right)]$, where the expectation is taken over the PPP $\Phi_{\rm D}(t)$ and the trajectories. This is essentially the average rate experienced by the typical UE at time $t$ across different network and trajectory realizations.

\noindent\emph{Session rate}: This metric is defined as the average received rate by the typical UE at each session of duration $T$. Mathematically speaking, we have
\begin{equation} \label{Eq:DefSR}
{\rm SR}(T) = \frac{1}{T}\int_0^T R(t)\,{\rm d}t.
\end{equation}

\section{Point Process of Interferers} \label{sec:Density}
In this section, we characterize the temporal evolution of the point process of interferers for all the mobility models described in Section \ref{subsec2:MobilityModels}. We start our analysis by first considering the UIM in the following lemma.
\vspace{-0.2cm}
\begin{lemma} \label{lem:NoExclusion}
Let $\Phi$ be a homogeneous PPP with density $\lambda_0$. If all the points of $\Phi$ are independently displaced based on the four mobility models mentioned in \ref{subsec2:MobilityModels}, then the displaced points at every time $t$ form another homogeneous PPP $\Psi$ with the same density $\lambda_0$.
\end{lemma}
\vspace{-0.2cm}
\begin{IEEEproof}
Based on the displacement theorem \cite{B_Haenggi_Stochastic_2012}, we need to argue that the displaced distances of DBSs at every time $t$ are i.i.d. and also independent of their original locations in $\Phi$. For the first two mobility models, this is clear from our model constructions. For the RW and RWP models, note that different flights are independent from the locations of DBSs and the overall displaced distance at time $t$ is a function of these flights. Hence, the overall displacement of DBSs are independent from their locations in the original PPP $\Phi$. Therefore, $\Psi$ is distributed as a homogeneous PPP with density $\lambda_0$. This completes the proof.
\end{IEEEproof}
In the UIM, the serving DBS and all the interfering DBSs are displaced in uniformly random directions based on our different mobility models. Thus, we can infer from Lemma \ref{lem:NoExclusion} that the network of all DBSs at any time $t$ will remain a homogeneous PPP with density $\lambda_0$. As a result, the network of {\em interfering} DBSs will be distributed as an inhomogeneous PPP in the DBS plane with density
\begin{equation} \label{LambdaServiceModel1}
\lambda(t; u_\nbx, u_0)=\left\{\begin{matrix}
\lambda_0 & u_\nbx > u_0(t)\\ 
0 & u_\nbx \leq u_0(t)
\end{matrix}.\right.
\end{equation}
Note that although the serving distance $u_0(t)$ varies over time, its distribution does not change. In the UDM, it is clear from our construction that $\Phi_{\rm D}'(0)$ is an inhomogeneous PPP with density given by \eqref{LambdaServiceModel1}, which introduces an \emph {exclusion zone}, $\ncalX = b({\bf o'}, u_0)$, for the interfering DBSs in the DBS plane, where $b({\bf o}, r)$ is a disc of radius $r$ centered at ${\bf o}$. Using displacement theorem, we observe that $\Phi_{\rm D}'(t)$ remains an inhomogeneous PPP for the UDM as well. In the next lemma, we provide a mathematical characterization of $\Phi_{\rm D}'(t)$ in the UDM at every time $t$.
\begin{lemma} \label{lem:MainDensity}
\emph {(Point Process of Interferers).} Consider the UDM with an i.i.d. mobility model. Let $\nbx(0)$ and $\nbx(t)$ denote the initial location and the location at time $t$ of an interfering DBS, respectively. Define $L(t) = \|\nbx(0) - \nbx(t)\|$ as the net displacement of this interfering DBS until time $t$ and denote its cumulative distribution function (cdf) and pdf by $F_L(l; t)$ and $f_L(l; t)$, respectively. Then the network of interfering DBSs will be an inhomogeneous PPP with density
\begin{align} \label{MainLambda}
\lambda(t; u_\nbx, u_0) = \lambda_0
  \begin{cases}
     1 & u_0 + vt \leq u_\nbx\\
     \beta(t, u_\nbx, u_0) & |u_0 - vt| \leq u_\nbx \leq u_0 + vt\\
     \beta(t, u_\nbx, u_0){\bf 1}\left(t>\frac{u_0}{v}\right) & 0 \leq u_\nbx \leq |u_0 - vt|
  \end{cases},
\end{align}
where ${\bf 1}(.)$ is the indicator function and
\begin{align} \label{MainBeta}
\beta(t, u_\nbx, u_0) = 1 - F_L(u_0 - u_\nbx; t) - \int_{|u_\nbx - u_0|}^{\min\{vt, u_\nbx + u_0\}} f_L(l; t)\frac{1}{\pi}\cos^{-1}\left(\frac{l^2 + u_\nbx^2 - u_0^2}{2lu_\nbx}\right)\,{\rm d}l.
\end{align}
\end{lemma}
\begin{IEEEproof}
See Appendix \ref{app:Lemma2}.
\end{IEEEproof}
\begin{remark} \label{Remark1}
We observe the following directly from Lemma \ref{lem:MainDensity}: (i) \eqref{MainLambda} is continuous at boundaries, i.e., at $u_\nbx = |u_0 \pm vt|$, (ii) as $u_0 \to 0$, we get $\beta(t, u_\nbx, u_0) \to 1$, i.e., the network of interfering DBSs becomes homogeneous, and (iii) as  $u_0 \to 0$ or $t \to 0$, \eqref{LambdaServiceModel1} and \eqref{MainLambda} become identical.
\end{remark}

The density derived in Lemma \ref{lem:MainDensity} is valid for all i.i.d. mobility models where DBSs move with the same constant speed. Hence, for mobility models in this paper, we only need to characterize the distribution of the net displacement of each DBS at every time $t$. For the SL model, we provide this distribution, and thus, the density of the network of interferers, in the next corollary.
\begin{cor} \label{cor:DensityModel1}
\emph {(Point Process of Interferers for the SL Model).} When the interferers move based on the SL mobility model, the network of interferers will be an inhomogeneous PPP with density
\begin{align} \label{MainLambda1}
\lambda(t; u_\nbx, u_0) = \lambda_0
  \begin{cases}
     1 & u_0 + vt \leq u_\nbx\\
     \frac{1}{\pi}\cos^{-1}\left(\frac{u_0^2 - u_\nbx^2 - v^2 t^2}{2u_\nbx v t}\right) & |u_0 - vt| \leq u_\nbx \leq u_0 + vt\\
     {\bf 1}\left(t>\frac{u_0}{v}\right) & 0 \leq u_\nbx \leq |u_0 - vt|
  \end{cases}.
\end{align}
\end{cor}
\begin{IEEEproof}
In the SL mobility model, we have $L(t) = vt$, and thus, $f_L(l; t) = \delta(l - vt)$ and $F_L(l; t) = {\bf 1}(l - vt)$, where $\delta(.)$ is the Dirac delta function. Hence, $\beta(t, u_\nbx, u_0)$ can be written as
\begin{align} \label{SubBeta1}
\beta(t, u_\nbx, u_0) = 1 - {\bf 1}(u_0 - u_\nbx - vt) - \int_{|u_\nbx - u_0|}^{\min\{vt, u_\nbx + u_0\}} \delta(l - vt)\frac{1}{\pi}\cos^{-1}\left(\frac{l^2 + u_\nbx^2 - u_0^2}{2lu_\nbx}\right)\,{\rm d}l.
\end{align}
We need to consider two cases:
\begin{itemize}
\item $0 \leq u_\nbx \leq |u_0 - vt|$\\
In this case, since $\lambda(t; u_\nbx, u_0) = 0$ when $u_0 \geq vt$, we need to only evaluate $\beta(t, u_\nbx, u_0)$ when $u_0 < vt$. This gives $0\leq u_\nbx \leq vt - u_0$, which makes the integral term in \eqref{SubBeta1} zero. Since $u_0 - u_\nbx \leq u_0 + u_\nbx \leq vt$, the term for the indicator function in \eqref{SubBeta1} will become zero as well. Hence, $\beta(t, u_\nbx, u_0) = {\bf 1}\left(t>\frac{u_0}{v}\right)$.
\item $|u_0 - vt| \leq u_\nbx \leq u_0 + vt$\\
In this case, we have the triangle inequality for the triple $(u_\nbx, u_0, vt)$, which yields $|u_0 - u_\nbx| \leq vt \leq u_0 + u_\nbx$. Hence,
\begin{align*}
\beta(t, u_\nbx, u_0) &= 1 - \frac{1}{\pi}\cos^{-1}\left(\frac{v^2 t^2 + u_\nbx^2 - u_0^2}{2vt u_\nbx}\right) = \frac{1}{\pi}\cos^{-1}\left(\frac{u_0^2 - u_\nbx^2 - v^2 t^2}{2u_\nbx vt}\right).
\end{align*}
\end{itemize}
This completes the proof.
\end{IEEEproof}

\begin{figure}[t!]
    \centering
    \includegraphics[width=0.55\columnwidth]{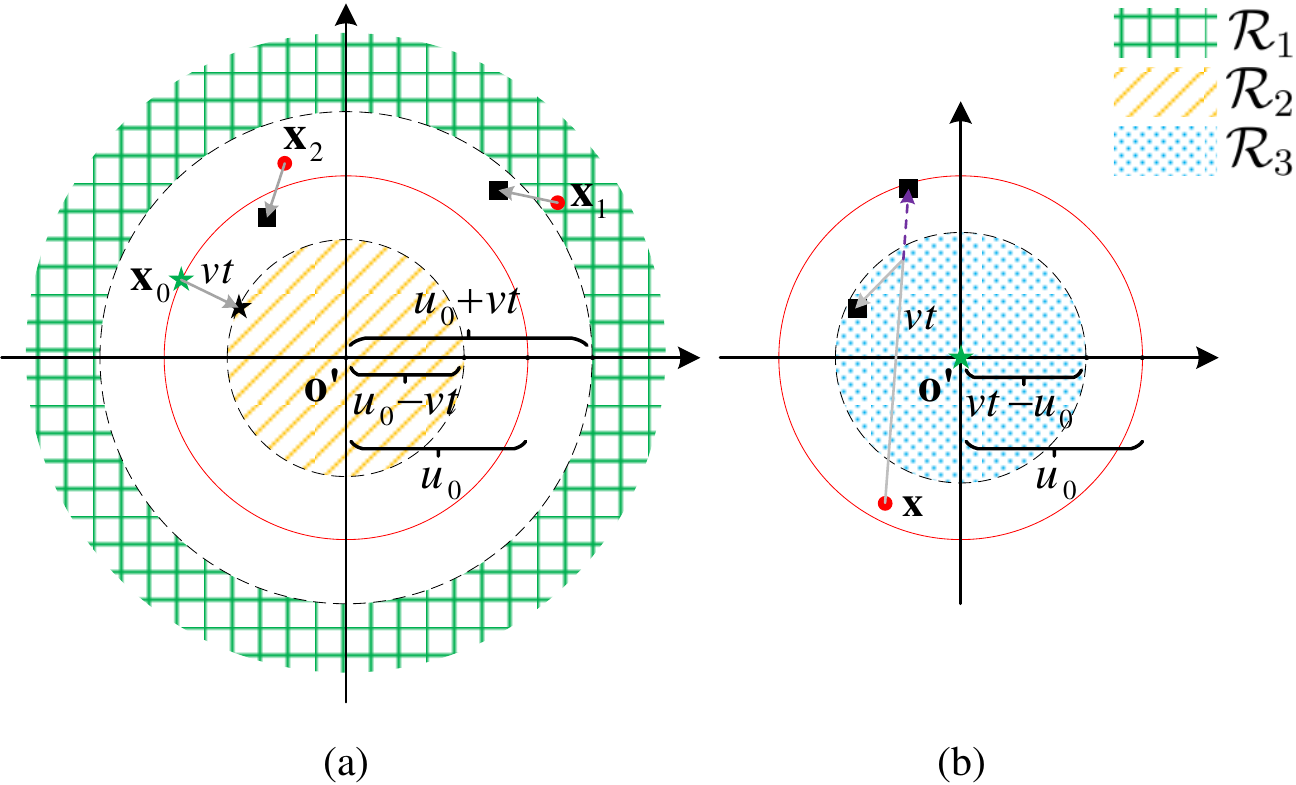}
    \vspace{-0.5cm}
    \caption{Network density in different regions for the SL mobility model. Green and black stars, red circles, and black squares represent serving DBS, displaced serving DBS, interfering DBSs, and displaced interfering DBSs, respectively. (a) Serving DBS is moving towards $\nbo'$, and (b) serving DBS is hovering at $\nbo'$.}
    \vspace{-0.8cm}
    \label{Fig:Remark2}
\end{figure}

\begin{remark} \label{rem:RegionsCompare}
We can make sense of \eqref{MainLambda1} intuitively by considering the exclusion zone $\ncalX$ and different regions illustrated in Fig. \ref{Fig:Remark2} as follows: (i) $\ncalR_1=\{u_\mathbf{x} \geq u_0 + vt\}$: No DBS that was initially in $\ncalR_1$ can enter $\ncalX$ within time $t$, and thus, $\lambda(t; u_\mathbf{x}, u_0) = \lambda_0$, (ii) $\ncalR_2=\{0 \leq u_\mathbf{x} \leq u_0 - vt | vt \leq u_0\}$: The serving DBS is in motion towards $\mathbf{o}'$ and since there is no interfering DBS in $\ncalR_2$, we have $\lambda(t; u_\mathbf{x}, u_0) = 0$, and (iii) $\ncalR_3=\{0 \leq u_\mathbf{x} \leq vt - u_0 | vt > u_0\}$: In this region, we can use the same method introduced in the proof of Lemma \ref{lem:MainDensity} as follows. We first calculate the density contributed by points that would have originally fallen in $\ncalX$ and then subtract it from $\lambda_0$ to get the density of the network of interfering DBSs. Since all the DBSs that are initially inside $\ncalX$ will leave $\ncalR_3$ after the displacement of $vt$, the density contributed by $\ncalX$ in $\ncalR_3$ is $0$, which gives $\lambda(t; u_\mathbf{x}, u_0) = \lambda_0$. However, for non-linear i.i.d. mobility models, such as RW or RWP, where direction changes are allowed during flights, there is a possibility that some of the DBSs that are initially inside $\ncalX$ do not leave $\ncalR_3$ after the displacement of $vt$. This fact is highlighted in Fig. \ref{Fig:Remark2} (b), where a DBS that is initially at $\nbx$ could possibly fall into $\ncalR_3$ after a displacement of $vt$ with a non-linear i.i.d. mobility model. Hence, the impact of $\ncalX$ is not zero in this case, which implies that the density of interfering DBSs is less than $\lambda_0$ in $\ncalR_3$.
\end{remark}
\vspace{-0.2cm}
It is in fact possible to make a more formal statement about the comparison of SL mobility model with the other models, which is done next.
\vspace{-0.2cm}
\begin{theorem} \label{thm:lowerbound}
The expected number of interferers in the disc $\ncalB = b(\nbo', u_0 + vt)$ at any time $t$ is maximized over the space of i.i.d. mobility models (including the ones where drones follow curved trajectories) when interferers follow the SL mobility model.
\end{theorem}
\vspace{-0.2cm}
\begin{IEEEproof}
See Appendix \ref{app:Theorem1}.
\end{IEEEproof}
\vspace{-0.2cm}
\begin{remark}
Theorem \ref{thm:lowerbound} demonstrates that the average number of interferers in any neighborhood of the typical UE is higher in the SL mobility model compared to the other i.i.d. mobility models. Consequently, the average received rate at the typical UE under the SL mobility model is lower compared to the other i.i.d. mobility models.
\end{remark}
\vspace{-0.2cm}
In the RS mobility model, the net displacement of DBSs until time $t$ is $L(t)=\min\{vt, R\}$, where $R$ is a random variable that determines the displaced distances of DBSs. The following corollary gives the network density for the RS mobility model.
\begin{cor} \label{cor:DensityModel2}
\emph {(Point Process of Interferers for the RS Model).} When interfering DBSs move based on the RS mobility model, the network of interfering DBSs will be an inhomogeneous PPP with the same density as given in Lemma \ref{lem:MainDensity}. Furthermore, the function $\beta(t, u_\nbx, u_0)$ can be written as
\begin{align} \label{MainBeta2}
 \scalebox{0.88}{$ \beta(t, u_\nbx, u_0) = F_R(u_\nbx - u_0) + \left( 1 - F_R(r) \right)\frac{1}{\pi}\cos^{-1}\left(\frac{u_0^2 - u_\nbx^2 - r^2}{2u_\nbx r}\right) + \int_{|u_\nbx - u_0|}^{r} f_R(l)\frac{1}{\pi}\cos^{-1}\left(\frac{u_0^2 - u_\nbx^2 - l^2}{2u_\nbx l}\right)\,{\rm d}l,$}
\end{align}
where $r = \min\{vt, u_\nbx + u_0\}$.
\end{cor}
\begin{IEEEproof}
Starting with $L(t)=\min\{vt, R\}$, we can write the cdf and pdf of $L(t)$ as
\begin{align*}
F_L(l; t) &= \nbbP[\min\{vt, R\} \leq l] = 1 - \nbbP[vt > l, R>l] = F_R(l){\bf 1}(vt - l) + {\bf 1}(l - vt),\\
f_L(l; t) &= (1 - F_R(vt))\delta(vt - l) + f_R(l){\bf 1}(vt - l).
\end{align*}
Similar to the proof of Corollary \ref{cor:DensityModel1}, we consider two cases to derive $\beta(t, u_\nbx, u_0)$ as follows.
\begin{itemize}
\item $0 \leq u_\nbx \leq |u_0 - vt|$.
\begin{align*}
\beta(t, u_\nbx, u_0) = 1 - F_R(u_0 - u_\nbx) - \int_{|u_\nbx - u_0|}^{u_\nbx + u_0} f_R(l)\frac{1}{\pi}\cos^{-1}\left(\frac{l^2 + u_\nbx^2 - u_0^2}{2lu_\nbx}\right)\,{\rm d}l.
\end{align*}
\item $|u_0 - vt| \leq u_\nbx \leq u_0 + vt$.
\begin{align*}
\scalebox{0.85}{$ \beta(t, u_\nbx, u_0) = 1 - F_R(u_0 - u_\nbx) - (1 - F_R(vt))\frac{1}{\pi}\cos^{-1}\left(\frac{v^2 t^2 + u_\nbx^2 - u_0^2}{2vt u_\nbx}\right) - \int_{|u_\nbx - u_0|}^{vt} f_R(l)\frac{1}{\pi}\cos^{-1}\left(\frac{l^2 + u_\nbx^2 - u_0^2}{2lu_\nbx}\right)\,{\rm d}l.$}
\end{align*}
\end{itemize}
Combining these equations into a single one, we end up with \eqref{MainBeta2} and the proof is complete.
\end{IEEEproof}
Note that the SL mobility model is a special case of the RS mobility model when $R \to \infty$, which means that the DBSs never stop. Mathematically speaking, evaluating $\beta(t, u_\nbx, u_0)$ in Corollary \ref{cor:DensityModel2} with $F_R(l)=0$ and $f_R(l)=\delta(l-\infty)=0$ for $l<\infty$, we end up with \eqref{MainLambda1}.

\vspace{-0.1cm}
\section{RW and RWP Mobility Models} \label{sec:RW}
In the previous section, we characterized the point process of interferers for both service models, which required the distribution of the net displacement of each DBS as a function of time. As it was shown in Corollaries \ref{cor:DensityModel1} and \ref{cor:DensityModel2}, this distribution can be easily derived for the SL and RS mobility models. However, characterization of this distribution for the RW and RWP mobility models is not straightforward and is the main focus of this section. Note that the distributional results provided here are novel and may be useful in their own right.

\begin{figure}
	\centering
	\includegraphics[width=0.35\columnwidth]{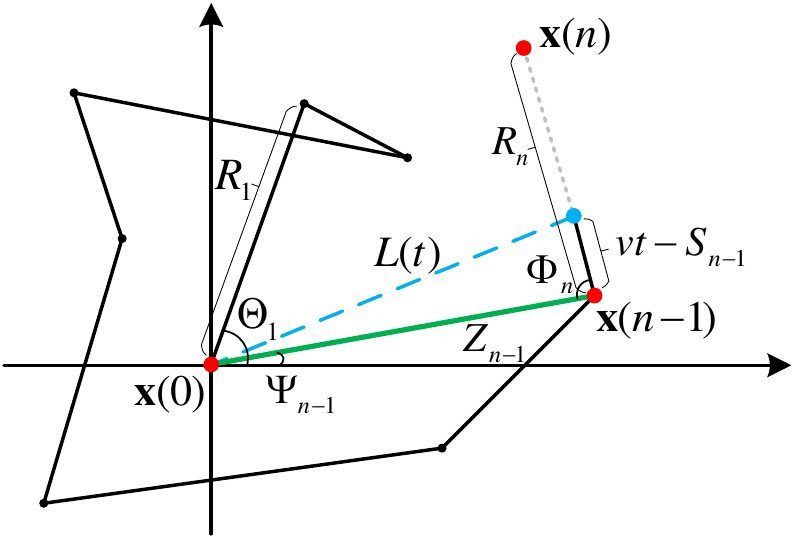}
	\vspace{-0.5cm}
	\caption{A realization of the RW mobility model.}
	\vspace{-0.8cm}
	\label{fig:RW}
\end{figure}

As stated in Definition \ref{def:MobilityModels}, we assume that the direction of the $i$-th movement of a DBS is $\Theta_i \sim U[0, 2\pi)$, which is selected independently of the other DBSs, and the flights $R_i$ are i.i.d. with cdf and pdf of $F_R(.)$ and $f_R(.)$, respectively. Fig. \ref{fig:RW} shows an example of the RW mobility model. Let $\nbx(0)$ and $\nbx(n-1)$ be the initial location of a DBS and its location after $(n-1)$ flights, respectively. Define $Z_{n-1}$ and $\Psi_{n-1}$ as the net displacement of a DBS between $\nbx(0)$ and $\nbx(n-1)$ and the angle between the $x$-axis and the line connecting $\nbx(0)$ and $\nbx(n-1)$, respectively. Furthermore, we define $S_{n-1}$ and $L(t)$ as the total distance traveled from $\nbx(0)$ to $\nbx(n-1)$ and the net displacement of a DBS until time $t$, respectively. In order to properly find the distribution of $L(t)$, we need the distributional characteristics of $S_{n-1}$, $Z_{n-1}$, and $\Psi_{n-1}$, which is done next.

From Fig. \ref{fig:RW}, we can derive the equations for $S_{n-1}$, $Z_{n-1}$, and $\Psi_{n-1}$ as
\vspace{-0.1cm}
\begin{align}\label{RW_Theta}
\scalebox{0.85}{$ S_{n-1} \!=\! \sum_{i=1}^{n-1} \! R_i $}, \hspace{0.45cm} \scalebox{0.85}{$ Z_{n-1} \!=\! \sqrt{\!\left(\sum_{i=1}^{n-1} \! R_i \cos(\Theta_i)\right)^2 \!\! + \! \left(\sum_{i=1}^{n-1} \! R_i \sin(\Theta_i)\right)^2}$}, \hspace{0.45cm} \scalebox{0.85}{$ \Psi_{n-1} \!=\! \tan^{-1}\!\left(\!\frac{\sum_{i=1}^{n-1} R_i \sin(\Theta_i)}{\sum_{i=1}^{n-1} R_i \cos(\Theta_i)}\!\right)$}.
\end{align}
\vspace{-0.1cm}
In the next lemma, we derive the distribution of $\Psi_n$.
\begin{lemma} \label{lem:RW_Psi_Uniform}
When interfering DBSs move based on the RW mobility model, then irrespective of the flight distribution $R$, the random variable $\Psi_n$ is distributed uniformly in $[0, 2\pi)$. 
\end{lemma}
\vspace{-0.1cm}
\begin{IEEEproof}
See Appendix \ref{app:Lemma3}. 
\end{IEEEproof}
Using the same methodology as in the proof of Lemma \ref{lem:RW_Psi_Uniform}, one can also find the distribution of $Z_n$. However, it turns out that the $(n-1)$-fold integral that arises in the derivation of the pdf of $Z_n$ does not have a closed-form solution in general. Nevertheless, when the flights are i.i.d. Rayleigh random variables, we can derive the distribution of $Z_n$ in closed form.
\vspace{-0.1cm}
\begin{lemma} \label{lem:RW_Z_Rayleigh}
When interfering DBSs move based on the RW mobility model and the flights are Rayleigh distributed with parameter $\sigma$, then $Z_n$ is also Rayleigh distributed with parameter $\sigma\sqrt{n}$.
\end{lemma}
\vspace{-0.1cm}
\begin{IEEEproof}
See Appendix \ref{app:Lemma4}.
\end{IEEEproof}

Lemma \ref{lem:RW_Z_Rayleigh} establishes a useful result for the distribution of $Z_n$ when the flights are Rayleigh distributed. On the other hand, if the flights have a general non-Rayleigh distribution, we can find an asymptotic distribution for $Z_n$ as $n \to \infty$. We present this result in the next lemma.
\begin{lemma} \label{lem:RW_Z_Rayleigh_Asymptotic}
When interfering DBSs move based on the RW mobility model and the flights have general non-Rayleigh but i.i.d. distributions with a mean and variance of $\mu_R$ and $\sigma_R^2$, respectively, then $\frac{Z_n}{\sqrt{n}}$ will have a Rayleigh distribution with parameter $\sqrt{\frac{\mu_R^2+\sigma_R^2}{2}}$ as $n \to \infty$.
\end{lemma}
\begin{IEEEproof}
Define $X = \sum_{i=1}^n \frac{1}{\sqrt{n}}R_i\cos(\Theta_i) $ and $Y = \sum_{i=1}^n \frac{1}{\sqrt{n}} R_i\sin(\Theta_i)$. Since $\Theta_i$'s are i.i.d. with uniform distribution in $[0, 2\pi)$, the central limit theorem (CLT) asserts that as $n \to \infty$, $X$ and $Y$ will have Gaussian distributions. Since $R_i$ and $\Theta_i$ are independent of each other, the moments of $X$ can be computed as follows: $\nbbE[X] = \sum_{i=1}^n \frac{1}{\sqrt{n}}\nbbE[R_i] \nbbE[\cos(\Theta_i)] = 0$ and $\nbbE[X ^ 2] = \nbbE\left[\sum_{i=1}^n\sum_{j=1}^n \frac{1}{n}R_i R_j \cos(\Theta_i)\cos(\Theta_j)\right] = \nbbE\left[\sum_{i=1}^n \frac{1}{n}R_i ^ 2 \cos ^ 2(\Theta_i)\right] = \frac{\mu_R^2+\sigma_R^2}{2}$. Note that the same is also true for $Y$. Hence, $X \sim \ncalN(0, \frac{\mu_R^2+\sigma_R^2}{2})$ and $Y \sim \ncalN(0,  \frac{\mu_R^2+\sigma_R^2}{2})$. Now since $\nbbE[XY] = 0$, $X$ and $Y$ are uncorrelated, and thus, independent. Therefore, $\frac{Z_n}{\sqrt{n}} = \sqrt{X^2 + Y^2}$ is Rayleigh distributed with parameter $\sqrt{\frac{\mu_R^2+\sigma_R^2}{2}}$ as $n \to \infty$. This completes the proof.
\end{IEEEproof}
After characterizing $Z_n$ and $\Psi_n$, we need to find the distribution of $S_n$ as well. Recall that $S_n$ is the sum of $n$ i.i.d. random variables and for a general distribution of $R_i$, the distribution of $S_n$ is not known. There have been some works that establish the distribution of $S_n$ for different distributions of $R_i$ (see \cite{J_Nadarajah_Review_2008} for a comprehensive review). When $R_i$'s are exponentially distributed, the distribution of $S_n$ is known to be Erlang. However, the exact distribution of $Z_n$ is not known for exponentially distributed flights. When $R_i$'s are Rayleigh, the distribution of $S_n$ has been investigated in the literature \cite{J_Hu_Accurate_2005} and the result is given here without proof.
\begin{lemma} \label{lem:RW_Sum_Rayleigh}
The pdf of the sum of $n$ i.i.d. Rayleigh random variables with parameter $\sigma$ can be approximated as
\begin{align}\label{sumNRayleigh}
\scalebox{1.01}{$ f_{S_n}(s) \approx \frac{(\frac{s}{\sqrt{n}})^{2n - 1} {\rm e}^{-\frac{s^2}{2nb}}}{2^{n - 1}b^n (n - 1)!} - \frac{(\frac{s}{\sqrt{n}} - a_2)^{2n - 2} {\rm e}^{-\frac{a_1 (\frac{s}{\sqrt{n}} - a_2)^2}{2b}}}{2^{n - 1}b (\frac{b}{a_1})^n (n - 1)!}\, a_0 \left[ b(2s\sqrt{n} - a_2) - a_1 \frac{s}{\sqrt{n}}(\frac{s}{\sqrt{n}} - a_2)^2 \right],$}
\end{align}
where $b = \sigma^2 \frac{\sqrt[n]{(2n - 1)!!}}{n}$ and the constants $a_0$, $a_1$, and $a_2$ are derived numerically using a nonlinear curve-fitting least square method based on the trust region reflective algorithm \cite{J_Coleman_Interior_1996}.
\end{lemma}


In order to better understand the distributional properties of the RW mobility model, we also need to find the joint distribution of $S_n$ and $Z_n$ for a given number of flights $n$. Given $S_n$, the range of the values for $Z_n$ is upper bounded by the value of $S_n$, which suggests a dependency between $S_n$ and $Z_n$. In the next proposition, we characterize this joint distribution.
\begin{prop}\label{prop:JointSZ}
The joint pdf of $S_n$ and $Z_n$ for a given number of flights $n \geq 2$ can be written as a $2(n-1)$-fold integral given as
\begin{align}\label{JointSZ}
f_{S_n, Z_n}(s, z) = \frac{4z}{(2\pi)^n}\underset{\ncalR}{\int\dots\int} \frac{f_R(s - J_{n}) \prod_{i=1}^{n-1} f_R(x_i)}{\sqrt{\left( z^2 - ( s - J_{n} - K_{n} )^2 \right)\left( ( s - J_{n} + K_{n} )^2 - z^2 \right)}}\, {\rm d}{\boldsymbol x} \, {\rm d}{\boldsymbol \xi},
\end{align}
where $J_{n} = \sum_{i=1}^{n-1} x_i$, $K_{n} = \sqrt{\sum_{i, j=1}^{n-1}x_i x_j \cos(\xi_i - \xi_j)}$, and $\ncalR$ is defined as the region where $0 \leq x_i < \infty$, $0 \leq \xi_i < 2\pi$ for $1 \leq i \leq n-1$, and $|s - J_{n} - K_{n}| \leq z \leq s - J_{n} + K_{n}$.
\end{prop}
\begin{IEEEproof}
See Appendix \ref{app:Proposition1}.
\end{IEEEproof}
Note that for $n = 1$, the random variables $S_n$ and $Z_n$ will become identical and equal to $R_1$, and thus, their joint distribution will be the same as the distribution of $R_1$, i.e., $f_R(.)$. For $n = 2$, the result of Proposition \ref{prop:JointSZ} can be further simplified, which is given in the next corollary.
\vspace{-0.3cm}
\begin{cor}
The joint pdf between $S_2$ and $Z_2$ can be written as
\begin{align}\label{JointS2Z2}
f_{S_2, Z_2}(s, z) = \frac{2z}{\pi\sqrt{s^2 - z^2}}\int_{\frac{s-z}{2}}^{\frac{s+z}{2}} \frac{f_R(x)f_R(s - x)  }{\sqrt{z^2 - (2x - s)^2}}\, {\rm d}x,
\end{align}
when $s > z$ and zero otherwise.
\end{cor}
\vspace{-0.6cm}
\begin{remark} \label{rem:independentSZ}
As $n$ gets larger, the dependency between $S_n$ and $Z_n$ will become less significant, and thus, we can ignore it for large enough $n$. Hence, we approximate the joint pdf of $S_n$ and $Z_n$ by assuming them to be independent for $n \geq 3$, thus giving $f_{S_n, Z_n}(s, z) \approx f_{S_n}(s)f_{Z_n}(z)$. Note that $n \geq 3$ is large enough for this result to be reasonably accurate.
\end{remark}
\vspace{-0.2cm}

Using the results that we have derived for the RW mobility model so far, we can now compute the distribution of $L(t)$, i.e., the distance between the location of a DBS at time $t$ and its original location at time $t=0$. Fig. \ref{fig:RW} shows a DBS that is flying in its $n$-th flight. Note that the distance traveled by the DBS until time $t$ is $vt$, and thus, the residual distance in the $n$-th flight will be $vt - S_{n-1}$. We now state the main result in the next proposition.
\vspace{-0.1cm}
\begin{prop}\label{prop:RWDistributionLt}
When interfering DBSs move based on the RW mobility model, the cdf and pdf of $L(t)$ are given as
\begin{align}
\hspace{-0.2cm}F_L(l; t) =\, &\scalebox{0.95}{$ (1 \!-\! F_R(vt)){\bf 1}(l \!-\! vt)$} + \sum_{n=2}^\infty \int_{vt - l}^{vt} \int_0^{l - (vt - s)} \scalebox{0.95}{$f_{S_{n - 1}, Z_{n - 1}}(s, z) (1 \!-\! F_R(vt \!-\! s))\, {\rm d}z \, {\rm d}s $} \,+ \nonumber\\
&\hspace{-1.1cm}\sum_{n=2}^\infty \int_{\frac{vt - l}{2}}^{vt} \int_{|l - (vt - s)|}^{\min\{s,\, l + (vt - s)\}}\scalebox{1}{$  f_{S_{n - 1}, Z_{n - 1}}(s, z)(1 \!-\! F_R(vt \!-\! s)) \frac{1}{\pi}\cos^{-1}\left( \frac{z^2 + (vt - s)^2 - l^2}{2z(vt - s)} \right)\, {\rm d}z \, {\rm d}s,$}\label{RWcdfLt}\\
\hspace{-0.2cm}f_L(l; t) =\, &\scalebox{0.95}{$ (1 \!-\! F_R(vt))\delta(l \!-\! vt)$} \!+\!  \frac{2l}{\pi}\sum_{n=2}^\infty \int_{\frac{vt - l}{2}}^{vt} \! \int_{|l - (vt - s)|}^{\min\{s,\, l + (vt - s)\}}\scalebox{0.92}{$ \!\!\!\!\!\frac{f_{S_{n - 1}, Z_{n - 1}}(s, z)(1 - F_R(vt - s))}{\sqrt{\left[ l^2 - ( z - (vt - s) )^2 \right]\left[ ( z + (vt - s) )^2 - l^2 \right]}} \, {\rm d}z \, {\rm d}s,$} \label{RWpdfLt}
\end{align}
when $l \leq vt$, respectively. Otherwise, we have $F_L(l; t) = 1$ and $f_L(l; t) = 0$.
\end{prop}
\begin{IEEEproof}
See Appendix \ref{app:Proposition2}.
\end{IEEEproof}
For $n=2$, the double integrals in \eqref{RWcdfLt} and \eqref{RWpdfLt} can be written as single integrals since $S_1 = Z_1 = R_1$. Mathematically speaking, when $l < vt$ we have
\begin{align}
F_L(l; t | n = 2) &= \int_{\frac{vt - l}{2}}^{\frac{vt + l}{2}} f_R(r)(1 - F_R(vt - r))\frac{1}{\pi}\cos^{-1}\left( \frac{r^2 + (vt - r)^2 - l^2}{2r(vt - r)} \right)\, {\rm d}r,\label{RWcdfLtn2}\\
f_L(l; t | n = 2) &= \frac{2l}{\pi\sqrt{v^2 t^2 - l^2}}\int_{\frac{vt - l}{2}}^{\frac{vt + l}{2}} \frac{f_R(r)(1 - F_R(vt - r))}{\sqrt{l^2 - ( 2r - vt )^2}}\, {\rm d}r.\label{RWpdfLtn2}
\end{align}
This result will be useful for our further approximations on the distribution of $L(t)$. Applying the results of Proposition \ref{prop:RWDistributionLt} to Lemma \ref{lem:MainDensity}, we can compute $\beta(t, u_\nbx, u_0)$ in \eqref{MainBeta} and get the density of the network of interfering DBSs for the RW mobility model.

Building on the RW mobility model, we now derive the distribution of $L(t)$ for the RWP mobility model as well. The RWP mobility model is defined by a sequence of quadruples at the $i$-th flight period: two waypoints $\nbp_{i-1}$ and $\nbp_i$ as the starting and destination waypoints, respectively, one transition length $R_i$, and one waiting time $T_i$ at the destination waypoint. In an infinite network, we can observe that the only difference between the RWP and the RW mobility models is the inclusion of the random variable $T_i$ as the waiting time at the end of each flight. Assuming that the random variables $\{T_i\}$ are i.i.d. with cdf and pdf of $F_T(.)$ and $f_T(.)$, respectively, and also independent from $R_i$, the following proposition extends the results of Proposition \ref{prop:RWDistributionLt} to the RWP mobility model. Before stating the main result, we define $W_n$ as the aggregate waiting time until the end of the $n$-th flight, i.e., $W_n = \sum_{i=0}^n T_i$. Note that $W_n$ will be independent from $S_n$ and $Z_n$. Moreover, we assume that $W_0 = T_0 \neq 0$, which implies that there is an initial random waiting time before the DBSs start to move.
\begin{prop}\label{prop:RWPDistributionLt}
When interfering DBSs move based on the RWP mobility model, the cdf and pdf of $L(t)$ are given as
\begin{align}
F_L(l; t) =\, &\int_{t - \frac{l}{v}}^\infty \left(1 - F_R(vt - vw)\right)f_T(w)\, {\rm d}w \,+ \nonumber\\
&\sum_{n=2}^\infty \int_{0}^{vt} \int_0^{\min\{s, l\}} f_{S_{n - 1}, Z_{n - 1}}(s, z) \left( F_{W_{n-2}}(t - \frac{s}{v}) - F_{W_{n-1}}(t - \frac{s}{v}) \right)\, {\rm d}z \, {\rm d}s \,+ \nonumber\\
&\sum_{n=2}^\infty \int_{vt - l}^{vt} \int_0^{l - (vt - y)}\!\!\! \int_0^{\frac{y}{v}} f_{W_{n-1}}(w) f_{S_{n - 1}, Z_{n - 1}}(y - vw, z) (1 - F_R(vt - y))\, {\rm d}w \, {\rm d}z \, {\rm d}y \,+ \nonumber\\
&\sum_{n=2}^\infty \int_{\frac{vt - l}{2}}^{vt} \int_{|l - (vt - y)|}^{\min\{y,\, l + (vt - y)\}} \!\!\!\!\int_0^{\frac{y}{v}} f_{W_{n-1}}(w) f_{S_{n - 1}, Z_{n - 1}}(y - vw, z) (1 - F_R(vt - y)) \,\times\nonumber\\
&\hspace{4.8cm}\frac{1}{\pi}\cos^{-1}\left( \frac{z^2 + (vt - y)^2 - l^2}{2z(vt - y)} \right)\, {\rm d}w \, {\rm d}z \, {\rm d}y,\label{RWPcdfLt}\\
f_L(l; t) =\, &\frac{1}{v}\left(1 - F_R(l)\right)f_T\left(t - \frac{l}{v}\right) + \sum_{n=2}^\infty \int_{l}^{vt} \scalebox{0.87}{$f_{S_{n - 1}, Z_{n - 1}}(s, l) \left( F_{W_{n-2}}(t - \frac{s}{v}) - F_{W_{n-1}}(t - \frac{s}{v}) \right) \, {\rm d}s $} \,+ \nonumber\\
&\frac{2l}{\pi}\sum_{n=2}^\infty \int_{\frac{vt - l}{2}}^{vt} \int_{|l - (vt - y)|}^{\min\{y,\, l + (vt - y)\}}\!\!\!\! \int_0^{\frac{y}{v}} \scalebox{1.03}{$ \frac{f_{W_{n-1}}(w) f_{S_{n - 1}, Z_{n - 1}}(y - vw, z)(1 - F_R(vt - y))}{\sqrt{\left[ l^2 - ( z - (vt - y) )^2 \right]\left[ ( z + (vt - y) )^2 - l^2 \right]}} \, {\rm d}w \, {\rm d}z \, {\rm d}y, $} \label{RWPpdfLt}
\end{align}
when $l \leq vt$, respectively. Otherwise, we have $F_L(l; t) = 1$ and $f_L(l; t) = 0$.
\end{prop}
\begin{IEEEproof}
See Appendix \ref{app:Proposition3}.
\end{IEEEproof}
As in the RW mobility model, the results can be further simplified for $n=2$ and $l < vt$ as
\begin{align}
F_L(l; t | n \!=\! 2) \!= \!&\int_{0}^{l}\!\! \scalebox{0.85}{$ f_R(r)\!\left(F_T(t \!-\! \frac{r}{v}) \!-\! F_{W_1}(t - \frac{r}{v})\right) {\rm d}r$} +\! \int_{t-\frac{l}{v}}^{t}\!\int_{0}^{vt - vw}\!\!\!\!\!\!\!\!\!\!\!\! \scalebox{0.85}{$f_R(r)f_{W_1}(w)(1 \!-\! F_R(vt\! -\! vw\! -\! r))\, {\rm d}r\, {\rm d}w$} \, + \nonumber\\
&\hspace{-2.7cm}\int_0^{t-\frac{l}{v}}\int_{\frac{vt - vw - l}{2}}^{\frac{vt - vw + l}{2}} \scalebox{1.02}{$  f_R(r)f_{W_1}(w)(1 - F_R(vt - vw - r)) \frac{1}{\pi}\cos^{-1}\left( \frac{r^2 + (vt - vw - r)^2 - l^2}{2r(vt - vw - r)} \right)\, {\rm d}r\, {\rm d}w,$}\label{RWPcdfLtn2}\\
f_L(l; t | n = 2) = \, &f_R(l)\left(F_T(t - \frac{l}{v}) - F_{W_1}(t - \frac{l}{v})\right)\, + \nonumber\\
&\frac{2l}{\pi}\int_0^{t-\frac{l}{v}}\int_{\frac{vt - vw - l}{2}}^{\frac{vt - vw + l}{2}} \frac{f_R(r)f_{W_1}(w)(1 - F_R(vt - vw - r))}{\sqrt{\left[(vt - vw)^2 - l^2\right]\left[l^2 - ( 2r - (vt - vw) )^2\right]}} \, {\rm d}r\, {\rm d}w.\label{RWPpdfLtn2}
\end{align}
Similar to the RW scenario, we can apply the distribution of $L(t)$ derived in Proposition \ref{prop:RWPDistributionLt} to Lemma \ref{lem:MainDensity} and compute the density of the network of interferers for the RWP mobility model.

\section{Average and Session Rates} \label{sec:Rate}
In this section, we compute the ${\rm SIR}$-based metrics defined in Section \ref{subsec4:Metrics} for the typical UE in the network. Equipped with the density of the network of interfering DBSs for each mobility model under both the UIM and the UDM, we first derive the average received rate by the typical UE in the UDM. The following theorem provides this result.
\vspace{-0.2cm}
\begin{theorem} \label{theo:ASE}
In the UDM and using any i.i.d. mobility model described in the previous sections, the average received rate by the typical UE at time $t$ is given as
\begin{align} \label{MainRates}
R(t) &= \int_0^\infty \int_0^\infty \frac{2\pi\lambda_0 u_0 {\rm e}^{-\pi \lambda_0 u_0^2}}{1+\gamma}\sum_{k=0}^{m_0-1} \frac{(-s)^k}{k!}\frac{\partial^k}{\partial s^k}\mathcal{L}_{I(t)}(s\bigr\rvert\mathbf{x}_0(t))\biggr|_{s=m_0\gamma r_0^\alpha(t)}\,{\rm d}u_0\,{\rm d}\gamma,
\end{align}
where $\mathcal{L}_{I(t)}(s\bigr\rvert\mathbf{x}_0(t))$ is the conditional Laplace transform of interference given as
\begin{align} \label{LaplaceIt}
\mathcal{L}_{I(t)}(s\bigr\rvert\mathbf{x}_0(t)) &= \exp\left[-2\pi\int_{0}^\infty \scalebox{1.02}{$ u_\mathbf{x}(t) \lambda(t; u_\mathbf{x}, u_0) \Big(1-\big( 1 + \frac{s \left(u_\mathbf{x}^2(t) + h^2\right)^{-\alpha/2}}{m} \big)^{-m}\Big) \,{\rm d}u_\mathbf{x}(t)$}\right].
\end{align}
\end{theorem}
\begin{IEEEproof}
We start by writing the complementary cumulative distribution function (ccdf) of ${\rm SIR(t)}$ conditioned on $\mathbf{x}_0(t)$ as
\begin{align*}
\mathbb{P}\left[{\rm SIR}(t) \geq \gamma \bigr\rvert\mathbf{x}_0(t) \right] \!&=\! \mathbb{E}\left[\mathbb{P}\left[ h_0(t) \geq \gamma r_0^\alpha(t)I(t)\middle|\mathbf{x}_0(t), I(t) \right]\right] \overset{(a)}{=} \mathbb{E}\left[ \frac{\Gamma(m_0, m_0\gamma r_0^\alpha(t)I(t))}{\Gamma(m_0)}\middle|\mathbf{x}_0(t) \right]\nonumber\\
&\hspace{-3.1cm}\overset{(b)}{=} \mathbb{E}\left[ \sum_{k=0}^{m_0-1} \frac{(m_0\gamma r_0^\alpha(t)I(t))^k}{k!} {\rm e}^{-m_0\gamma r_0^\alpha(t)I(t)} \middle|\mathbf{x}_0(t) \right] = \sum_{k=0}^{m_0-1} \frac{(-s)^k}{k!} \frac{\partial^k}{\partial s^k}\mathcal{L}_{I(t)}(s\bigr\rvert\mathbf{x}_0(t))\biggr|_{s=m_0\gamma r_0^\alpha(t)},
\end{align*}
where in $(a)$ the Nakagami-$m$ fading assumption is used and in $(b)$ we used the definition of the incomplete gamma function for integer values of $m_0$. The conditional Laplace transform of interference at time $t$ can be computed as
\begin{align*}
\mathcal{L}_{I(t)}(s\bigr\rvert\mathbf{x}_0(t)) &= \mathbb{E}\left[{\rm e}^{-sI(t)}\bigr\rvert\mathbf{x}_0(t)\right] = \mathbb{E}\left[\exp\left[-s\sum_{\mathbf{x}(t) \in \Phi_{\rm D}'(t)} h_\mathbf{x}(t) r_\mathbf{x}(t)^{-\alpha}\right] \middle|u_0(t) \right] \nonumber\\
&\hspace{-2.5cm}\overset{(a)}{=} \mathbb{E}\!\!\left[ \prod_{\mathbf{x}(t) \in \Phi_{\rm D}'(t)}\!\!\!\! \scalebox{0.9}{$\left( 1 + \frac{s r_\mathbf{x}(t)^{-\alpha}}{m} \right)^{\!-m} $} \middle|u_0(t)\right]\! \overset{(b)}{=} \!\exp\left[-2\pi\!\!\int_{0}^\infty \!\!\!\!\!\scalebox{0.9}{$ u_\mathbf{x}(t) \lambda(t; u_\mathbf{x}, u_0) \left(1\!-\!\left( 1 \!+\! \frac{s r_\mathbf{x}(t)^{\!-\alpha}}{m} \right)^{-m}\right) {\rm d}u_\mathbf{x}(t)$}\right],
\end{align*}
where (a) results from the moment generating function (MGF) of the gamma distribution and (b) follows from the probability generating functional (PGFL) of a general PPP. Now, the average rate at time $t$ can be written as $R(t) = \mathbb{E}\left[\log\left(1 + {\rm SIR}(t)\right)\right] = $
\begin{align*}
\int_0^\infty \log(1+\gamma)f_\Gamma(\gamma; t)\,{\rm d}\gamma = \int_0^\infty \int_0^\infty \frac{2\pi\lambda_0 u_0 {\rm e}^{-\pi \lambda_0 u_0^2}}{1+\gamma}\mathbb{P}\left[{\rm SIR}(t) \geq \gamma \bigr\rvert\mathbf{x}_0(t) \right]\,{\rm d}u_0\,{\rm d}\gamma,
\end{align*}
where $f_\Gamma(\gamma; t)$ is the pdf of ${\rm SIR}(t)$ and in the last equation, we used integration by parts and deconditioned the result on $u_0(t)$. Note that in the UDM, we have $u_0(t) = [u_0 - vt]^+$, where $[x]^+=x$ if $x\geq 0$ and $[x]^+=0$ otherwise. This completes the proof.
\end{IEEEproof}
For the UIM, since the density of the network of interfering DBSs is given in \eqref{LambdaServiceModel1}, the received rate by the typical UE will be given as in \eqref{MainRates} evaluated at $t=0$. Finally, the session rate at time $T$ is derived by integrating over the average rate with respect to $t$ as in \eqref{Eq:DefSR}.

\vspace{-0.1cm}
\section{Simulation Results} \label{sec:ITbounds}
In this section, we provide numerical simulations to verify our analytical results and provide several useful insights about the system-level performance of this network. We assume that the network density is $\lambda_0 = 10 ^ {-6}$ and DBSs move with a constant speed $v = 45~{\rm km/h}$ using one of the four mobility models defined in Section \ref{sec:SysMod}. We consider the low altitude platform (LAP) for the flight of DBSs and assume $h \in \{100, \, 200\}~{\rm m}$ as typical values for the height in LAP. Furthermore, we assume the path loss exponent is $\alpha = 3$. For the RS, RW, and RWP mobility models, we assume that the flight distances are distributed as i.i.d. Rayleigh random variables with mean $500~{\rm m}$. Moreover, in the RWP mobility model, the hovering times follow i.i.d. exponential random variables with mean $5~{\rm s}$. Note that the evaluations of the multiple integrals that occur in the analyses of the RW and RWP mobility models are carried out using the Monte Carlo (MC) integration method.

\begin{figure}
\centering
\includegraphics[width=1\columnwidth]{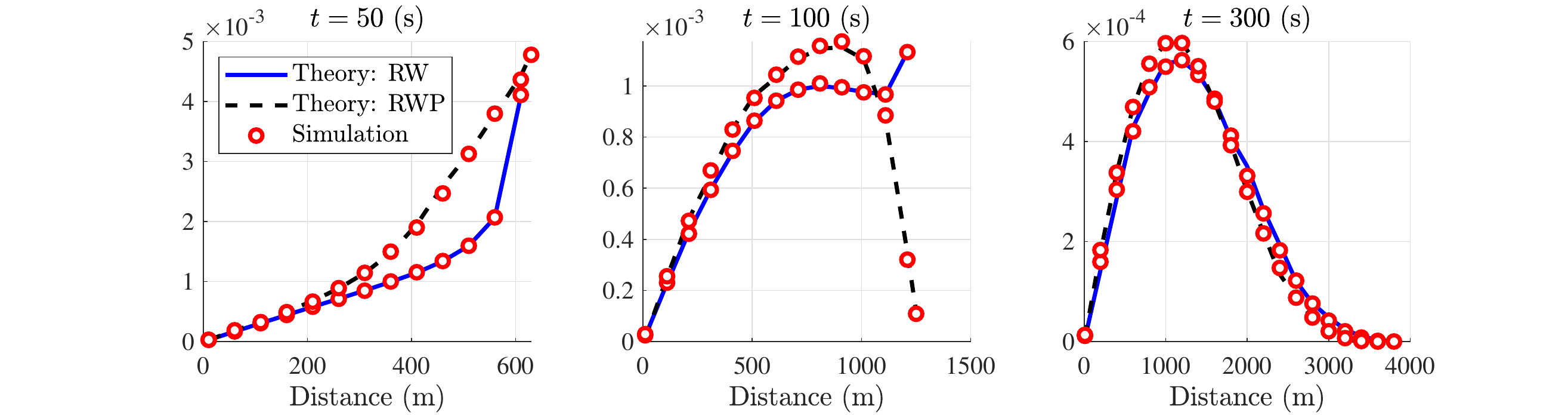}
\vspace{-1.1cm}
\caption{Distribution of $L(t)$ in the RW and RWP mobility models for $t \in \{50, 100, 300\}~{\rm m}$. As $t$ increases, the pdf of $L(t)$ in both the RW and RWP models converge to a Rayleigh distribution.}
\vspace{-0.7cm}
\label{fig:Sim1_Distributions}
\end{figure}

\vspace{-0.4cm}
\subsection{Distribution of $L(t)$ in RW and RWP}
Characterizing the distribution of $L(t)$ for a given time $t$ is essential for the analysis of the RW and RWP mobility models. We derived the exact cdf and pdf of $L(t)$ for the RW and RWP mobility models in Propositions \ref{prop:RWDistributionLt} and \ref{prop:RWPDistributionLt}, respectively. Fig. \ref{fig:Sim1_Distributions} shows the pdf of $L(t)$ for both the RW and RWP mobility models at $t \in \{50, 100, 300\}~{\rm s}$. Note that as $t \to \infty$, the distance traveled during the $n$-th flight will be much smaller than $Z_{n-1}$, and thus, we have $L(t) \approx Z_{n-1}$. Now, according to Lemma \ref{lem:RW_Z_Rayleigh_Asymptotic}, as $t \to \infty$, the distribution of $Z_{n-1}$ converges to a Rayleigh distribution. Hence, $L(t)$ is also Rayleigh distributed as $t \to \infty$. This trend can also be noticed in Fig. \ref{fig:Sim1_Distributions}.

\vspace{-0.4cm}
\subsection{Point Process of Interferers}
\begin{figure}
	\centering
	\includegraphics[width=1\columnwidth]{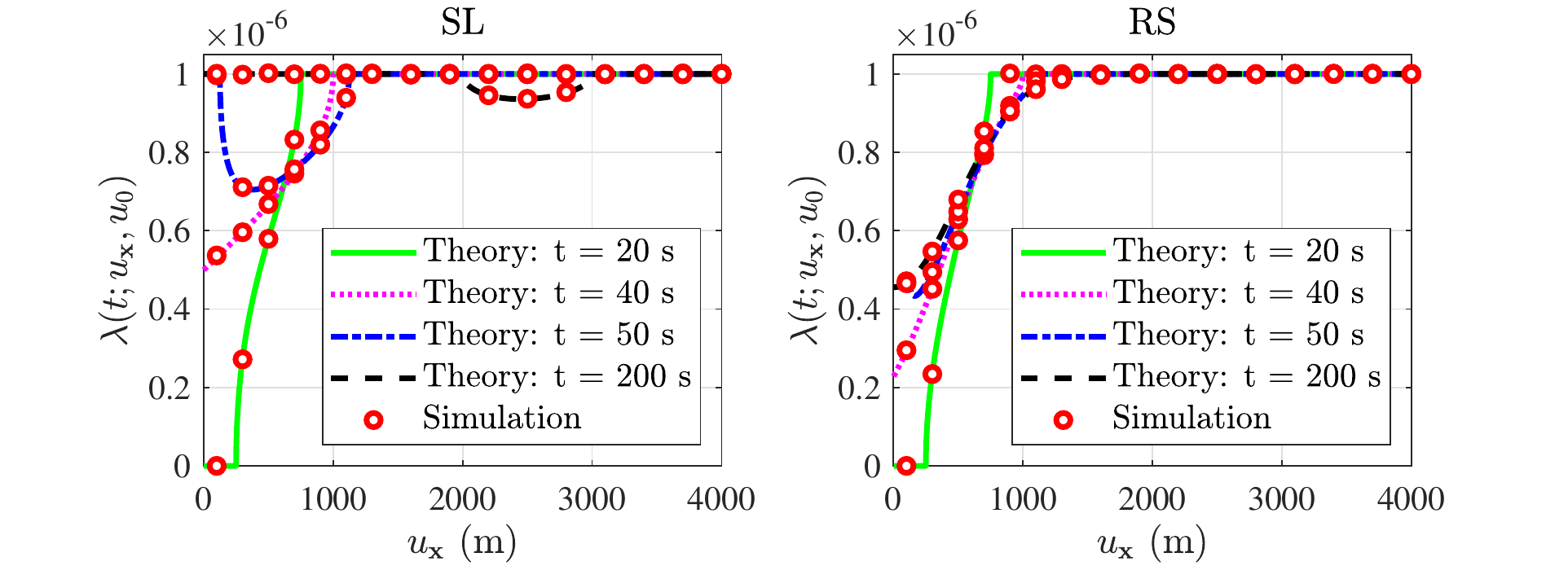}
	\vspace{-1.2cm}
	\caption{Density of the network of interfering DBSs for the UDM with the SL and RS mobility models. Serving distance is $u_0 = 500~{\rm m}$ and the density is given at $t \in \{20, 40, 50, 200\}~{\rm s}$.}
	\vspace{-0.4cm}
	\label{fig:Sim2_Densities12}
\end{figure}

\begin{figure}
	\centering
	\includegraphics[width=1\columnwidth]{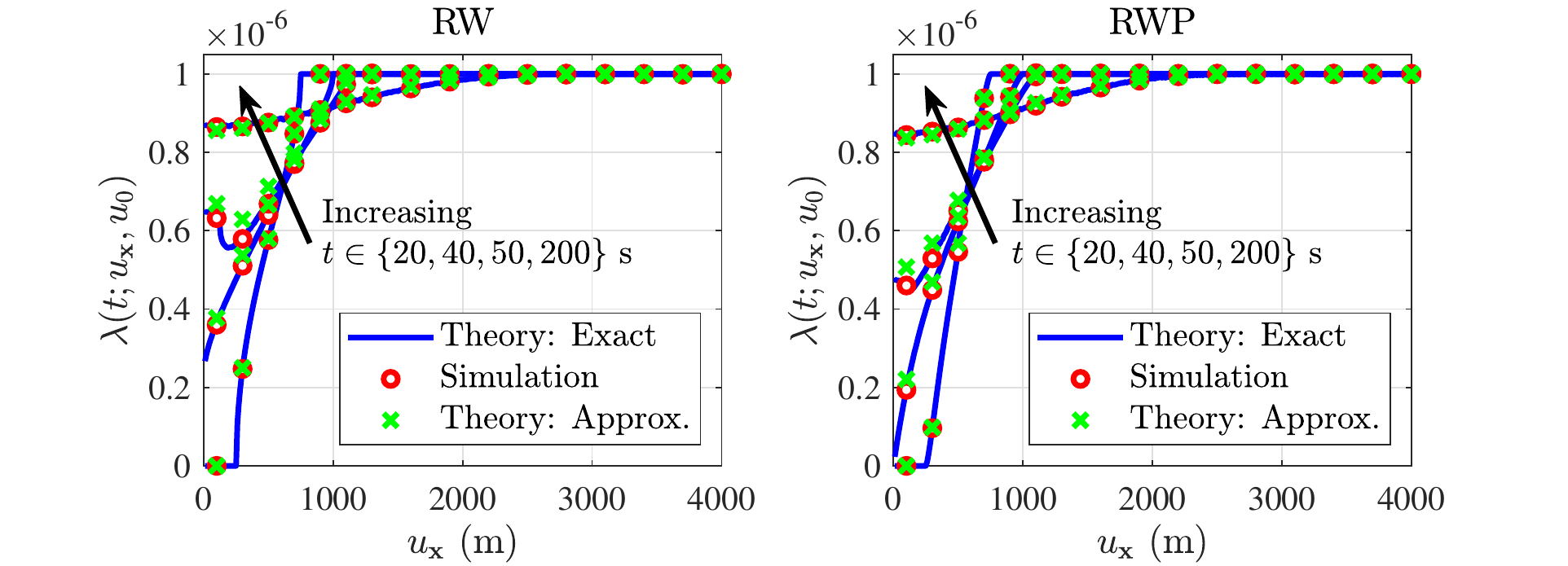}
	\vspace{-1.2cm}
	\caption{Density of the network of interfering DBSs for the UDM with the RW and RWP mobility models.
		Serving distance is $u_0 = 500~{\rm m}$ and the flights are Rayleigh distributed with mean $500~{\rm m}$.}
	\vspace{-0.7cm}
	\label{fig:Sim2_Densities34}
\end{figure}

In Figs. \ref{fig:Sim2_Densities12} and \ref{fig:Sim2_Densities34}, we plot the density of the network of interfering DBSs for all the mobility models considered in this paper at $t \in \{20, 40, 50, 200\}~{\rm s}$. We assume that the serving DBS follows the UDM and the exclusion zone radius is $u_0 = 500~{\rm m}$. In the SL mobility model, as also highlighted in Remark \ref{rem:RegionsCompare}, the density will be divided into two homogeneous parts and one bowl-shaped inhomogeneous part after $t = \frac{u_0}{v}$. Furthermore, the inhomogeneous part will become homogeneous as $t\to\infty$, which ultimately makes the point process of interferers homogeneous. This fact can also be directly inferred from Corollary \ref{cor:DensityModel1} by taking the limit of \eqref{MainLambda1} as $t \to \infty$ and $u_\nbx \to vt$. According to Fig. \ref{fig:Sim2_Densities34}, this homogenization happens for the RW and RWP mobility models as well. However, this is not the case in the RS mobility model, since the DBSs ``stop" moving after a period of time. Hence, the point process of interferers does not evolve with time, making it inhomogeneous for all time $t$. In Fig. \ref{fig:Sim2_Densities34}, based on Remark \ref{rem:independentSZ}, we also plot the density of the network of interfering DBSs for the RW and RWP mobility models by assuming that $S_n$ and $Z_n$ are independent for $n \geq 3$. Clearly, this approximation is fairly accurate.

\vspace{-0.4cm}
\subsection{Impact of Fading and Height}
\begin{figure}[t!]
	\centering
	\begin{minipage}{0.48\columnwidth}
		\centering
		\includegraphics[width=0.8\textwidth]{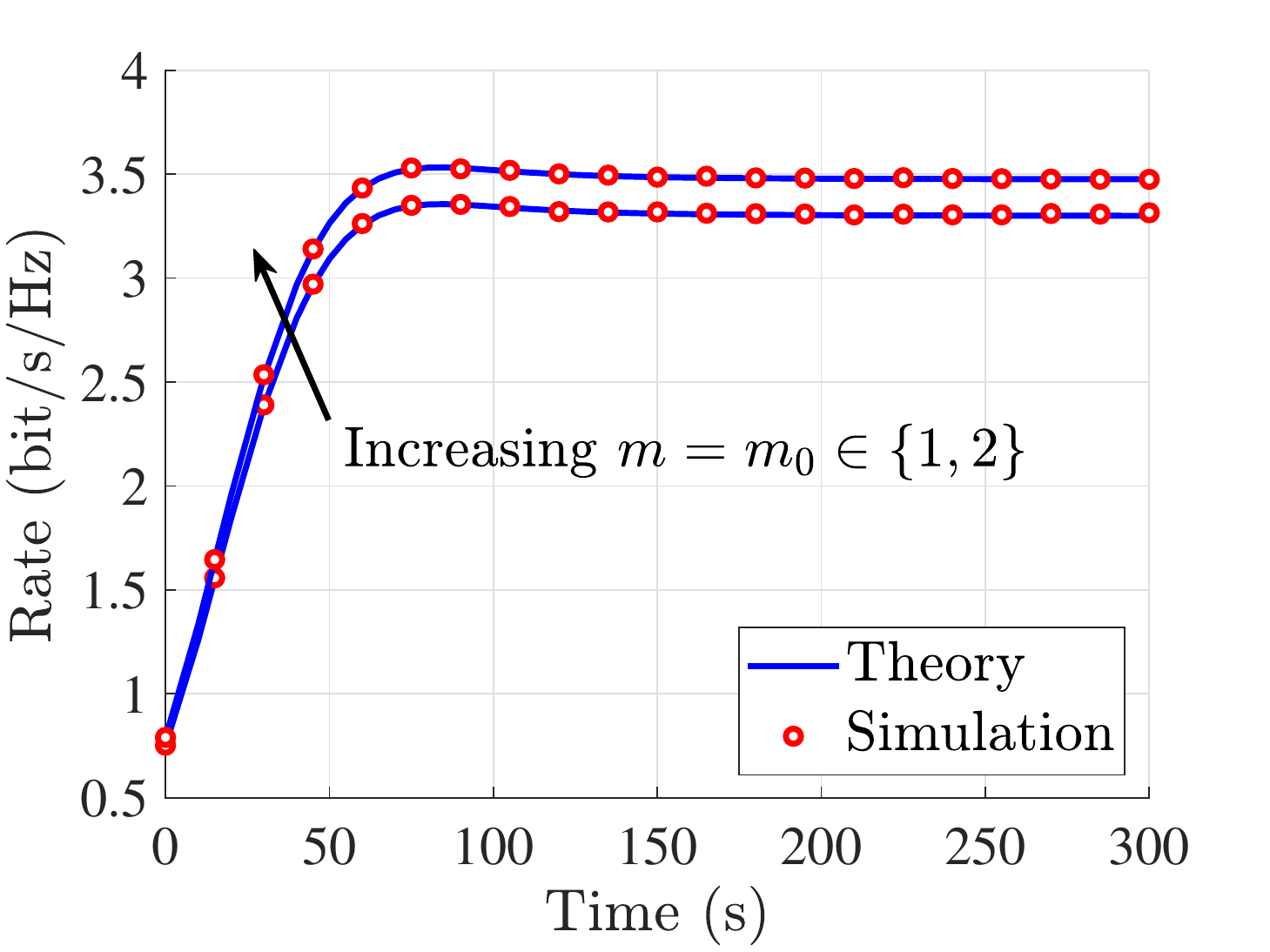}
		\vspace{-0.5cm}
		\caption{Comparison of the average rate for the UDM in the SL mobility model for different values of the Nakagami-$m$ fading parameter with $\alpha = 3$ and $h = 100~{\rm m}$.}
		\vspace{-0.6cm}
		\label{fig:Sim3_RatePlot_VaryFade}
	\end{minipage}\hfill
	\begin{minipage}{0.48\columnwidth}
		\centering
		\includegraphics[width=0.8\textwidth]{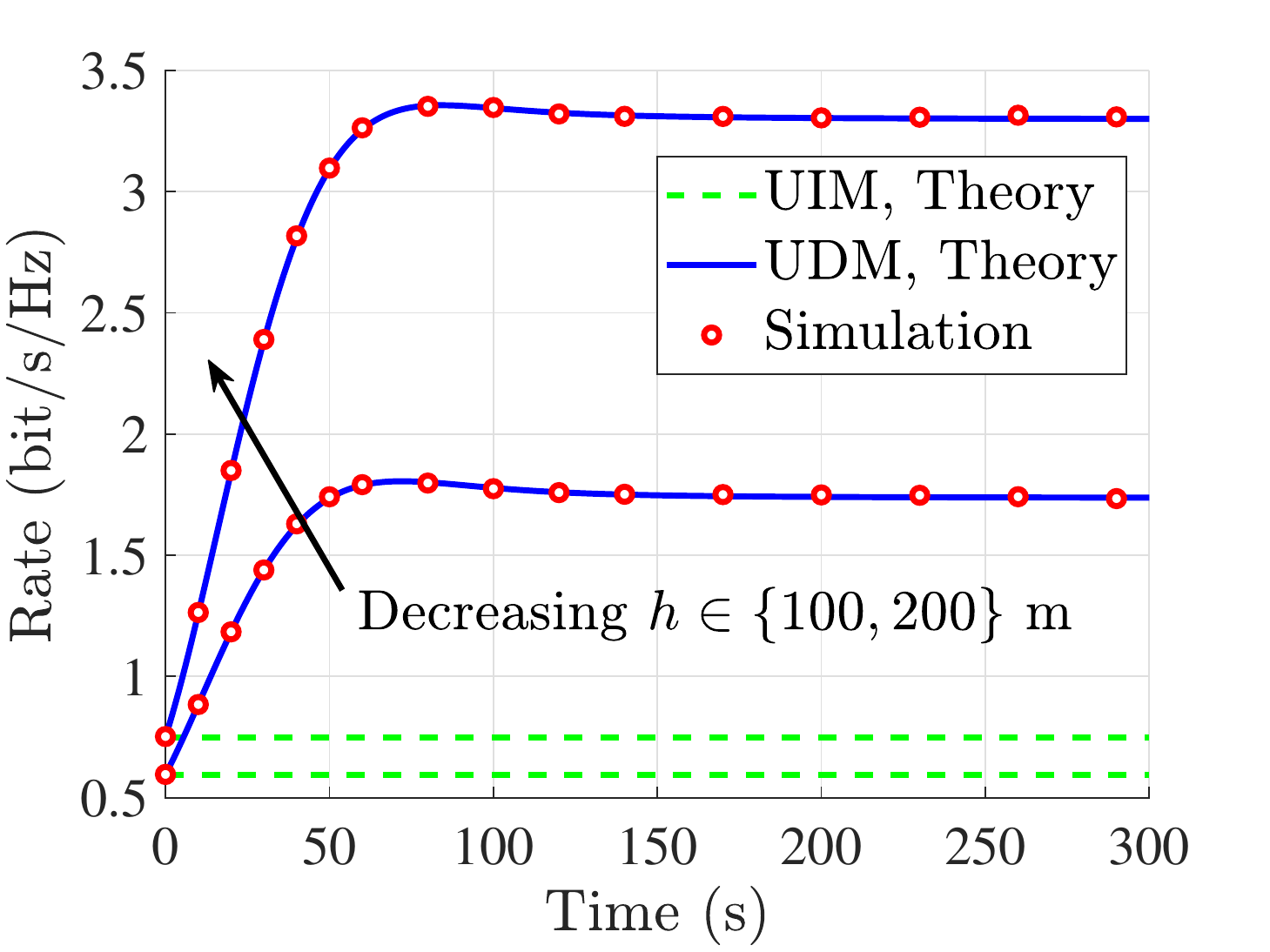}
		\vspace{-0.5cm}
		\caption{Comparison of the average rate for both the UIM and the UDM in the SL mobility model at different heights with $\alpha = 3$ and $m = m_0 = 1$.}
		\vspace{-0.6cm}
		\label{fig:Sim4_RatePlot_VaryHeight}
	\end{minipage}
\end{figure}

In order to show the effect of the Nakagami-$m$ fading parameter on the performance of the network, we plot the average rate in the UDM under the SL mobility model for $m = m_0 \in \{1, 2\}$ in Fig. \ref{fig:Sim3_RatePlot_VaryFade}. Since increasing $m$ and $m_0$ decreases the severity of the fading, the average rate will increase as well. Fig. \ref{fig:Sim4_RatePlot_VaryHeight} depicts the impact of DBS heights in the performance of the network for both service models under the SL mobility model. As is clear in this figure, the average rate increases as height decreases, which can also be observed directly from \eqref{MainRates} and \eqref{LaplaceIt}.

\vspace{-0.4cm}
\subsection{Impact of Mobility Models}
\begin{figure}[t!]
    \centering
    \begin{minipage}{0.48\columnwidth}
        \centering
        \includegraphics[width=0.8\textwidth]{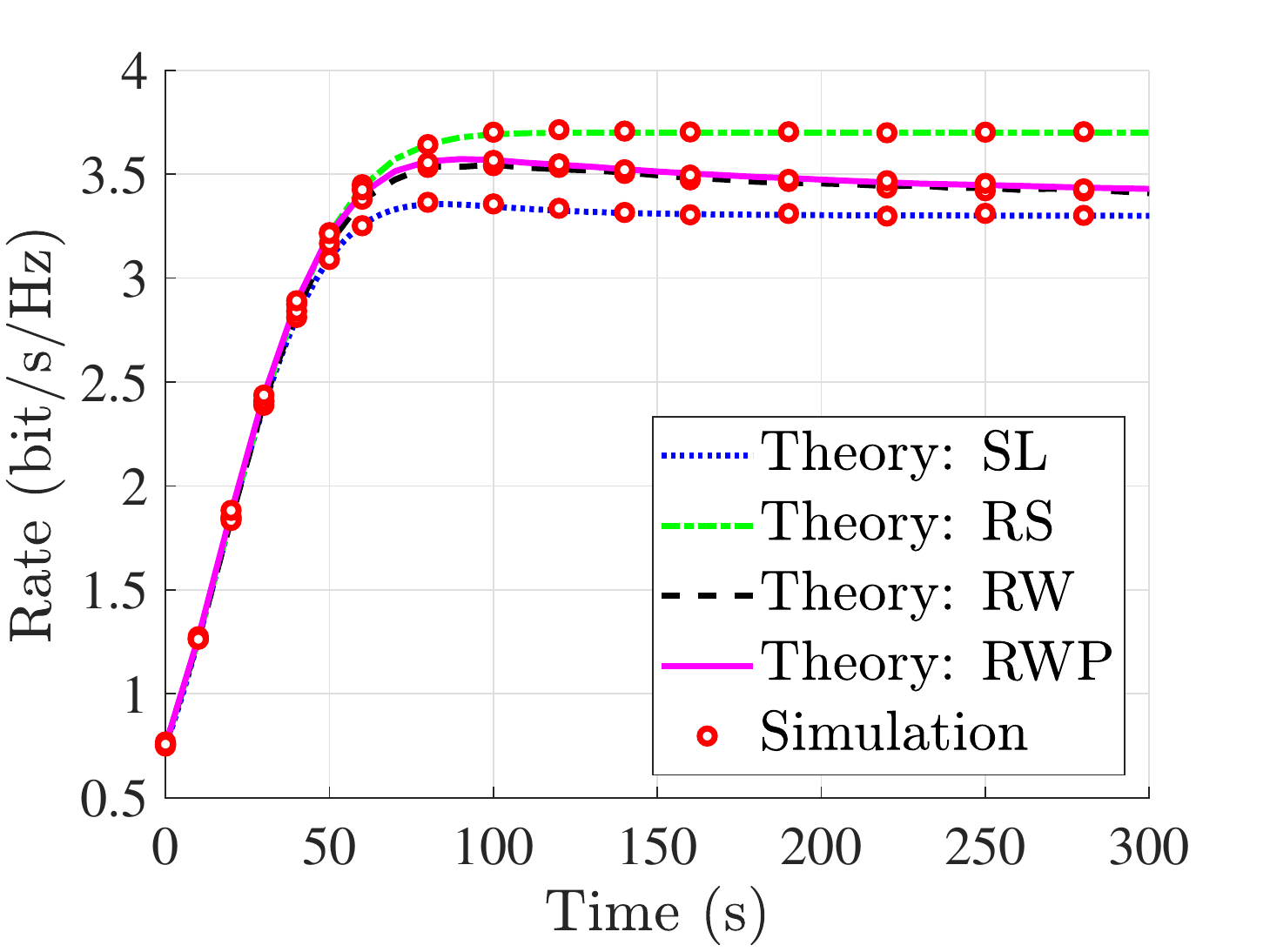}
        \vspace{-0.5cm}
        \caption{Comparison of the average rate for the UDM in different mobility models. Other parameters are $\alpha = 3$, $h = 100~{\rm m}$, and $m = m_0 = 1$.}
        \vspace{-0.8cm}
        \label{fig:Sim5_RatePlot_VaryMobilityModel}
    \end{minipage}\hfill
    \begin{minipage}{0.48\columnwidth}
        \centering
        \includegraphics[width=0.8\textwidth]{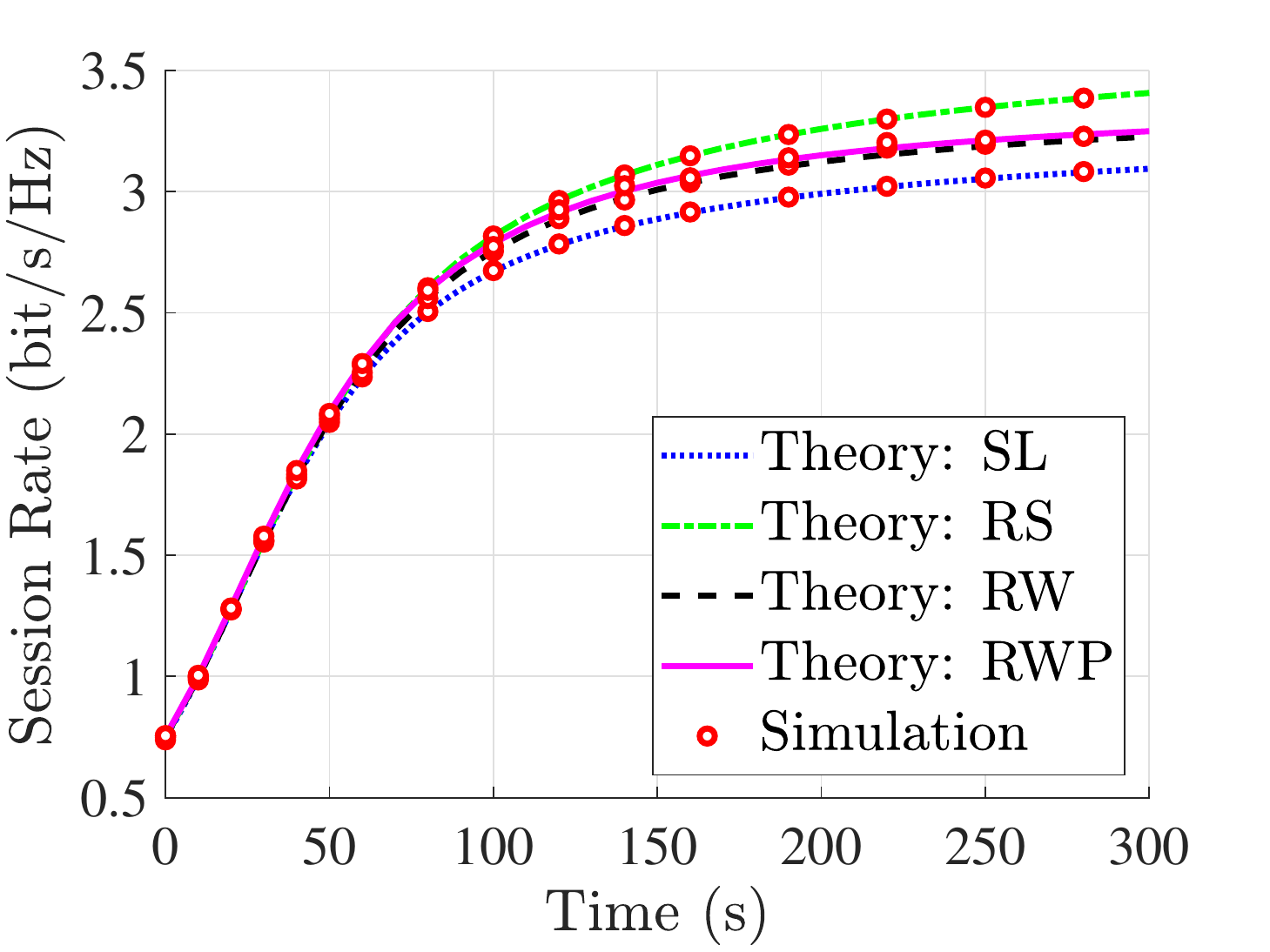}
        \vspace{-0.5cm}
        \caption{Comparison of the session rate for the UDM in different mobility models. Other parameters are $\alpha = 3$, $h = 100~{\rm m}$, and $m = m_0 = 1$.}
        \vspace{-0.8cm}
        \label{fig:Sim6_SessionRatePlot_VaryMobilityModel}
    \end{minipage}
\end{figure}

Under the UDM, we compare different mobility models studied in this paper in terms of the average and session rates in Figs. \ref{fig:Sim5_RatePlot_VaryMobilityModel} and \ref{fig:Sim6_SessionRatePlot_VaryMobilityModel}, respectively. As can be seen in these figures, the SL mobility model acts as a lower bound on the performance to these models. This result is essentially the one that we proved in Theorem \ref{thm:lowerbound} by demonstrating that the average number of interferers in the vicinity of the typical UE is higher in the SL mobility model than the other i.i.d. mobility models. Hence, the performance of the network under the SL mobility model will be worse than the other mobility models.

\vspace{-0.4cm}
\section{Conclusion}
In this paper, we presented an in-depth and unified analysis of a mobile drone cellular network operating at a constant height to serve the UEs on the ground. Specifically, we considered four mobility models for the DBSs, i.e., (i) SL, (ii) RS, (iii) RW, and (iv) RWP, and provided several fundamental distributional properties of these models. The use of the SL mobility model for drone networks was inspired by the simulation models used in the 3GPP studies of drone networks, while the others are useful canonical models (or their variants) that have been used extensively in wireless networks and provide a reasonable balance between realism and tractability. The serving DBS is selected based on the nearest-neighbor association policy and moves according to two service models: (i) based on the same mobility model as the interfering DBSs (UIM), and (ii) towards the typical UE at a constant height and keeps hovering above the location of the typical UE (UDM). We proposed a novel characterization of the point process of DBSs for both the UIM and the UDM, using which we analyzed the average received rate and the session rate at the typical UE. Borrowing tools from the calculus of variations, we mathematically showed that the SL mobility model acts as a lower bound on the system-level performance of our mobile drone network over the space of all \emph{i.i.d.} mobility models. To the best of our understanding, this is the first work that offers a unified analysis of canonical mobility models for a drone cellular network in an infinite plane and establishes meaningful connections between them. While this work offers many useful insights in the canonical settings, it will be useful to extend some of these results to more realistic mobility models, including the ones developed from actual mobility traces of drones (as and when they become available).

\vspace{-0.2cm}
\appendix
\vspace{-0.3cm}
\subsection{Proof of Lemma \ref{lem:MainDensity}} \label{app:Lemma2}
Since we have started with an inhomogeneous PPP with the density given as \eqref{LambdaServiceModel1} for $t=0$ in the UDM and the displacements are independent of each other, the network of interfering DBSs will also be an inhomogeneous PPP at every time $t$ due to displacement theorem \cite{B_Haenggi_Stochastic_2012}. According to Lemma \ref{lem:NoExclusion}, if there was no exclusion zone and DBSs moved independently of each other, as in our mobility models, the network of all DBSs (including the serving DBS) would have remained a homogeneous PPP with density $\lambda_0$. Taking $\ncalX$ into account, the resulting density of the network can be partitioned into two sets: (i) density contributed by $\ncalX$ (denoted as $\lambda_1(t; u_\nbx, u_0)$), i.e., due to the points that are initially inside $\ncalX$, and (ii) density of interferers (denoted as $\lambda(t; u_\nbx, u_0)$), i.e., due to the points that are initially outside $\ncalX$. Since the resulting network density is $\lambda_0$, we have $\lambda(t; u_\nbx, u_0) = \lambda_0 - \lambda_1(t; u_\nbx, u_0)$.
\begin{figure}
\centering
\includegraphics[width=0.35\columnwidth]{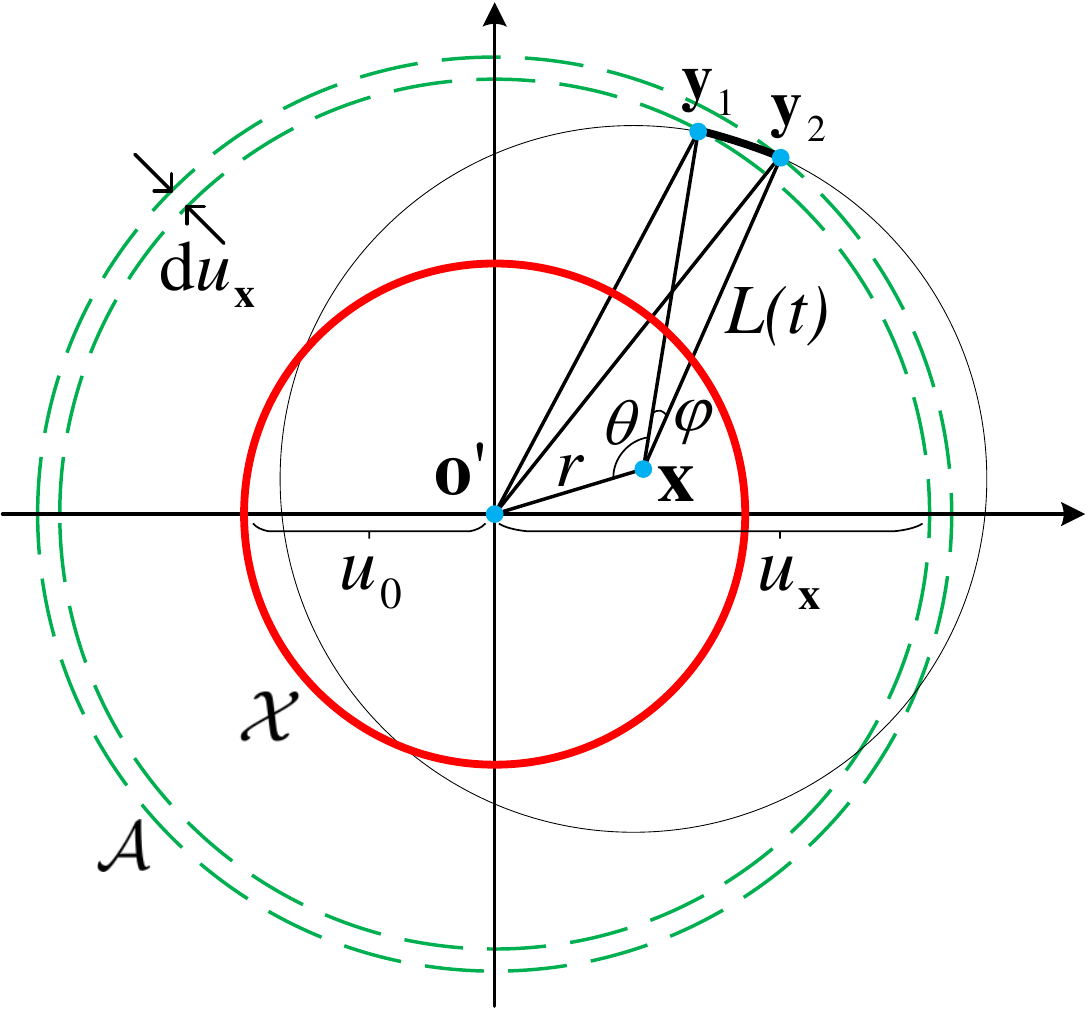}
\vspace{-0.5cm}
\caption{An illustration for the proof of Lemma \ref{lem:MainDensity}. The red circle and the green dotted circles indicate $\ncalX$ and $\ncalA$, respectively.}
\vspace{-0.7cm}
\label{fig:ProofTheorem1_1}
\end{figure}
Define $N(t)$ as the average number of points that are initially inside $\ncalX$ and after a displacement of $L(t)$ land on an infinitesimal annulus $\ncalA$ with an inner and outer radii of $u_\nbx$ and $u_\nbx + {\rm d}u_\nbx$, respectively. Since the density in this case is rotation invariant, it is sufficient to consider an annulus centered at $\nbo'$ for our analysis. By definition, we can write
\begin{align} \label{DefLambda}
\lambda_1(t; u_\nbx, u_0) &= \lim_{{\rm d}u_\nbx \to 0}\frac{N(t)}{2\pi u_\nbx {\rm d}u_\nbx},
\end{align}
where the denominator denotes the area of $\ncalA$. As shown in Fig. \ref{fig:ProofTheorem1_1}, let $\nbx$, $\nby_1$, and $\nby_2$ denote a randomly selected point inside $\ncalX$ with distance $r$ from ${\bf o'}$ and two intersection points of $b(\nbx, L(t))$ with $\ncalA$, respectively. Writing the cosine law in triangles ${{\bf o'} \nbx \nby_1}$ and ${{\bf o'} \nbx \nby_2}$, we have
\begin{align*}
u_\nbx^2 = L(t)^2 + r^2 - 2rL(t)\cos(\theta), \hspace{1cm}(u_\nbx + {\rm d}u_\nbx)^2 = L(t)^2 + r^2 - 2rL(t)\cos(\theta + \varphi),
\end{align*}
where $\theta = \angle{\bf o'}\nbx \nby_1$ and $\varphi = \angle \nby_1 \nbx \nby_2$. Eliminating $\theta$ in both of these equations gives
\begin{align*}
\cos(\varphi) &= \frac{1}{b^2}\Bigg[a^2 - a\left({\rm d}u_\nbx + \frac{({\rm d}u_\nbx)^2}{2u_\nbx}\right) + c\sqrt{c^2 + 2a\left({\rm d}u_\nbx + \frac{({\rm d}u_\nbx)^2}{2u_\nbx}\right) - \left({\rm d}u_\nbx + \frac{({\rm d}u_\nbx)^2}{2u_\nbx}\right)^2}\Bigg],
\end{align*}
where $a = \frac{L(t)^2 + r^2 - u_\nbx^2}{2u_\nbx}, b = \frac{2rL(t)}{2u_\nbx}, c = \sqrt{b^2 - a^2}$. Note that the probability that a DBS at $\nbx$ lands on $\ncalA$ after a displacement of $L(t)$ is $\frac{2\varphi}{2\pi}$. Hence, we write $N(t)$ by considering all such points $\nbx$ in $\ncalX$ as $N(t) = \nbbE\left[\int_0^{u_0} \frac{\varphi}{\pi} 2\pi r \lambda_0\,{\rm d}r\right]$, where the expectation is taken over $L(t)$. This gives
\begin{align}\label{EqLim1}
\lambda_1(t; u_\nbx, u_0) &= \frac{\lambda_0}{\pi}\nbbE\left[\int_0^{u_0} \frac{r}{u_\nbx} \lim_{{\rm d}u_\nbx \to 0}\frac{\varphi}{{\rm d}u_\nbx}\,{\rm d}r\right],
\end{align}
where we have changed the order of limit with the expectation and the integration due to the continuity of the integrand. Representing ${\rm d}u_\nbx$ with $x$ for simplicity, we compute the limit as
\begin{align}\label{EqLim2}
&\lim_{x \to 0}\frac{\varphi}{x} = \lim_{x \to 0}\frac{1}{x}\cos^{-1}\scalebox{1}{$ \Bigg(1 - \frac{1}{a^2 + c^2}\Bigg[c^2 + ax + \frac{a}{2 u_\nbx}x^2 - c\sqrt{c^2 + 2ax + \frac{a}{ u_\nbx }x^2 - \left( x + \frac{x^2}{2 u_\nbx } \right)^2}\Bigg]\Bigg)$}\nonumber\\
&\overset{(*)}= \!\!\sqrt{\frac{2}{a^2 + c^2}}\!\lim_{x \to 0}\scalebox{1.05}{$ \sqrt{\frac{(a^2 + c^2)\left(x^2 + \frac{1}{ u_\nbx }x^3 + \frac{1}{4 u_\nbx ^2}x^4\right)}{2c^2x^2}}$} = \frac{1}{c} = \frac{2 u_\nbx}{\sqrt{( u_\nbx ^2 \!-\! (L(t)\!-\!r)^2)((L(t)\!+\!r)^2 \!-\!  u_\nbx ^2)}},
\end{align}
where in $(*)$ we used the Taylor series expansion $\cos^{-1}(1-x) = \sqrt{2x} + \Theta(x^{3/2})$ as $x\to 0$, where $f(x) = \Theta(g(x))$ implies that $f(x)$ is asymptotically bounded by $g(x)$ both from above and below. Note that since the triangle inequality holds for the triple $(u_\nbx, r, L(t))$, the result in \eqref{EqLim2} is real and positive, as expected. Plugging \eqref{EqLim2} into \eqref{EqLim1}, we have
\begin{align} \label{Lambda1First}
\lambda_1(t; u_\nbx, u_0) = \frac{\lambda_0}{\pi}{\int_0^{\infty}}\int_{\ncalI_1}\frac{2r f_L(l;t)}{\sqrt{(u_\nbx^2 - (l-r)^2)((l+r)^2 - u_\nbx^2)}}\,{\rm d}r\,{\rm d}l,
\end{align}
where $\ncalI_1 = \left\{|l-u_\nbx| \leq r \leq l+u_\nbx\right\} \bigcap \left\{0 \leq r \leq u_0\right\}$. Since the net displacement of a DBS at time $t$ cannot exceed its aggregate traveled distance $vt$, we have $L(t) \leq vt$. Hence,
\begin{align*}
\lambda_1(t; u_\nbx, u_0) =&\, \frac{\lambda_0}{\pi}{\int_0^{vt}}\int_{\ncalI_1} f_L(l;t)\frac{2r}{\sqrt{(u_\nbx^2 - (l-r)^2)((l+r)^2 - u_\nbx^2)}}\,{\rm d}r\,{\rm d}l~+\nonumber\\
&\,\frac{\lambda_0}{\pi}(1-F_L(vt;t))\int_{\ncalI_2}\frac{2r}{\sqrt{(u_\nbx^2 - (vt-r)^2)((vt+r)^2 - u_\nbx^2)}}\,{\rm d}r,
\end{align*}
where $\ncalI_2 = \left\{|vt-u_\nbx| \leq r \leq vt+u_\nbx\right\} \bigcap \left\{0 \leq r \leq u_0\right\}$. Simplifying the last step requires tedious integrations and the details are skipped to maintain brevity. Finally, the density of the network of interfering DBSs is summarized as \eqref{MainLambda} and \eqref{MainBeta} in the lemma statement.
\hfill 
\IEEEQED
\vspace{-0.3cm}
\subsection{Proof of Theorem \ref{thm:lowerbound}} \label{app:Theorem1}
The intensity measure or the expected number of interfering DBSs in the Borel set $\ncalB = b(\nbo', u_0 + vt)$ is given as $\Lambda(\ncalB) = 2\pi\int_0^{u_0 + vt} u_\nbx \lambda(t; u_\nbx, u_0)\, {\rm d}u_\nbx$. Defining $\Lambda_1(\ncalB)$ and $\Lambda_2(\ncalB)$ as the intensity measures for the SL mobility model and an i.i.d. mobility model, respectively, we need to show that $\Lambda_1(\ncalB) \geq \Lambda_2(\ncalB)$. Note that we only prove this for $t \leq \frac{u_0}{v}$ and the proof for $t > \frac{u_0}{v}$ follows on the similar lines. Using \eqref{MainLambda} and \eqref{MainBeta}, we write $\Lambda_1(\ncalB) - \Lambda_2(\ncalB)$ as
\begin{align} \label{eq:lowerbound1}
\Lambda_1(\ncalB) \!-\! \Lambda_2(\ncalB) = 2\pi\lambda_0 \!\int_{u_0 - vt}^{u_0 + vt} \!\!\!u_\nbx\!\left[ \scalebox{0.9}{$  -g(vt, u_\nbx)\! +\! F_L(u_0 \!-\! u_\nbx; t)\, +\!\!$} \int_{|u_\nbx - u_0|}^{vt} \scalebox{0.9}{$ f_L(l; t) g(l, u_\nbx) \,{\rm d}l$} \right]\, {\rm d}u_\nbx,
\end{align}
where $g(l, u_\nbx) = \frac{1}{\pi}\cos^{-1}\left(\frac{l^2 + u_\nbx^2 - u_0^2}{2lu_\nbx}\right)$ and we denote the integrand by $\mathfrak{L}(u_\nbx, F_L(u_\nbx), f_L(u_\nbx))$, which is a functional with $u_\nbx$ being the independent variable. Note that we also have $F_L(0) = 0$ and $F_L(vt) = 1$ by definition. Using the Euler-Lagrange equation, we show that $\Lambda_1(\ncalB) - \Lambda_2(\ncalB)$ attains its minimum at zero. From the calculus of variations, we know that $F_L$ is a critical (extremum) function for the functional $\mathfrak{L}$, if it satisfies the Euler-Lagrange equation: $\frac{\partial \mathfrak{L}}{\partial F_L} - \frac{{\rm d}}{{\rm d}u_\nbx}\left( \frac{\partial \mathfrak{L}}{\partial f_L} \right) = 0$.
We compute the first term as $\frac{\partial \mathfrak{L}}{\partial F_L} = u_\nbx\mathbf{1}(u_0 - u_\nbx)$ and the second term as
\begin{align*}
\frac{{\rm d}}{{\rm d}u_\nbx}\left( \frac{\partial \mathfrak{L}}{\partial f_L} \right) = \frac{{\rm d}}{{\rm d}u_\nbx}\left( \int_{|u_\nbx - u_0|}^{vt} u_\nbx g(l, u_\nbx) \,{\rm d}l \right) \overset{(a)}{=} u_\nbx\mathbf{1}(u_0 - u_\nbx) + \int_{|u_\nbx - u_0|}^{vt} \frac{\partial}{\partial u_\nbx}\left(u_\nbx g(l, u_\nbx)\right) \,{\rm d}l,
\end{align*}
where in $(a)$ we used the Leibniz integral rule along with $g(|u_\nbx - u_0|, u_\nbx) = \mathbf{1}(u_0 - u_\nbx)$. Applying these derivatives to the Euler-Lagrange equation, we get $\int_{|u_\nbx - u_0|}^{vt} \frac{\partial}{\partial u_\nbx}\left(u_\nbx g(l, u_\nbx)\right) \,{\rm d}l = 0$, which does not include either $F_L$ or $f_L$. This means that the original function in \eqref{eq:lowerbound1} attains its extremum at the boundaries. We now evaluate $\mathfrak{L}(u_\nbx, F_L(u_\nbx), f_L(u_\nbx))$ at $u_\nbx = u_0 \pm vt$ as
\begin{align*}
u_\nbx &= u_0 - vt \longrightarrow \mathfrak{L} = -u_\nbx g(vt, u_0 - vt) + u_\nbx F_L(+vt) + 0 = 0\\
u_\nbx &= u_0 + vt \longrightarrow \mathfrak{L} = -u_\nbx g(vt, u_0 + vt) + u_\nbx F_L(-vt) + 0 = 0.
\end{align*}
Since both of the boundary values are $0$, we conclude that the minimum value in \eqref{eq:lowerbound1} is $0$.
\hfill 
\IEEEQED
\vspace{-0.3cm}
\subsection{Proof of Lemma \ref{lem:RW_Psi_Uniform}} \label{app:Lemma3}
In order to derive the distribution of $\Psi_n$, we introduce $(n - 1)$ auxiliary random variables $\Xi_i = \Theta_i$, $1 \leq i \leq n - 1$, and find the joint pdf of $n$ random variables $\Psi_n$ and $\Xi_i$, $1 \leq i \leq n - 1$. We then integrate out these auxiliary random variables to find the pdf of $\Psi_n$. We start by solving the system of $n$ equations (one equation in \eqref{RW_Theta} for the definition of $\Psi_n$ and $(n - 1)$ equations introduced by the auxiliary random variables) to derive $\Theta_i$'s in terms of $\Psi_n$ and $\Xi_i$'s. The result can be written as two sets of solutions as follows:
\begin{align*}
\scalebox{0.89}{$ 
{\rm Set~}1:
  \begin{cases}
	\Theta_i = \Xi_i, \hspace{0.5cm} i = 1, 2, \dots, n - 1\\
	\Theta_n = \Xi_{n_1} = \tan^{-1} \left( \frac{-\Delta\cos(\Psi_n) + \Delta'\sin(\Psi_n)}{\Delta\sin(\Psi_n) + \Delta'\cos(\Psi_n)} \right),
  \end{cases}
{\rm Set~}2:
  \begin{cases}
	\Theta_i = \Xi_i, \hspace{0.5cm} i = 1, 2, \dots, n - 1\\
	\Theta_n = \Xi_{n_2} = \tan^{-1} \left( \frac{-\Delta\cos(\Psi_n) - \Delta'\sin(\Psi_n)}{\Delta\sin(\Psi_n) - \Delta'\cos(\Psi_n)} \right),
  \end{cases}
  $}
\end{align*}
where $\Delta =  \sum_{i=1}^{n-1} R_i \sin(\Xi_i - \Psi_n)$ and $\Delta' = \sqrt{R_n^2 - \Delta^2}$. Note that $2\Psi_n = \Xi_{n_1} + \Xi_{n_2}$. Computing the determinant of the Jacobian matrix, we get $|\ncalJ| = |\frac{\partial \Theta_n}{\partial \Psi_n}|$ for both solution sets. Hence,
\begin{align*}
|\ncalJ| = \frac{\sum_{i=1}^n\sum_{j=1}^n R_i R_j \cos(\Theta_i - \Theta_j)}{\sum_{i=1}^n R_i R_n  \cos(\Theta_i - \Theta_n)} = 1 + \frac{\sum_{i=1}^n\sum_{j=1}^{n-1} R_i R_j \cos(\Theta_i - \Theta_j)}{\sum_{i=1}^n R_i R_n  \cos(\Theta_i - \Theta_n)}.
\end{align*}
By some algebraic manipulations, we find that $|\ncalJ|\bigr\rvert_{\Theta_n = \Xi_{n_1}} + |\ncalJ|\bigr\rvert_{\Theta_n = \Xi_{n_2}} = 2$. Now, since $\Theta_i \sim U[0, 2\pi)$, the joint distribution of $\Psi_n$ and $\Xi_i$'s can be written as
\begin{align*}
f_{\Psi_n, \boldsymbol {\Xi}}(\psi_n, \boldsymbol {\xi}) = f_{\boldsymbol {\Theta}}(\boldsymbol {\theta}) |\ncalJ|\bigr\rvert_{\Theta_n = \Xi_{n_1}} + f_{\boldsymbol {\Theta}}(\boldsymbol {\theta}) |\ncalJ|\bigr\rvert_{\Theta_n = \Xi_{n_2}} = 2\left(\frac{1}{2\pi}\right)^n,
\end{align*}
where the boldface letters represent vector random variables. Integrating out $\Xi_i$, $1 \leq i \leq n - 1$, the distribution of $\Psi_n$ is derived as $f_{\Psi_n}(\psi_n) = \frac{1}{\pi}$ for $\psi_n \in [-\frac{\pi}{2}, \frac{\pi}{2})$, due to the range of the $\tan^{-1}$ function. Finally, since the true range of $\Psi_n$ is $[-\pi, \pi)$, we conclude that $f_{\Psi_n}(\psi_n) = \frac{1}{2\pi}$ for $\psi_n \in [-\pi, \pi)$, and the proof is complete.
\hfill 
\IEEEQED
\vspace{-0.3cm}
\subsection{Proof of Lemma \ref{lem:RW_Z_Rayleigh}} \label{app:Lemma4}
We consider an RW of two flights $F_1$ and $F_2$, where $F_1$ and $F_2$ are two independent Rayleigh random variables with parameters $\sigma_1$ and $\sigma_2$, respectively. Let $F$ be the distance from the start of the first flight to the end of the second flight. We show that $F$ is Rayleigh distributed with parameter $\sigma = \sqrt{\sigma_1^2 + \sigma_2^2}$, which proves the lemma by induction. Since $F_1$ and $F_2$ are Rayleigh distributed, they can be written as $F_1 = \sqrt{X_1^2 + Y_1^2}$ and $F_2 = \sqrt{X_2^2 + Y_2^2}$, where $X_1, Y_1 \sim \ncalN(0, \sigma_1^2)$ are $X_2, Y_2 \sim \ncalN(0, \sigma_2^2)$ are independent Gaussian random variables. Note that due to the independence of $X_1$ and $X_2$, their sum will also be a Gaussian random variable. Therefore, $X = X_1 + X_2 \sim \ncalN(0, \sigma_1^2 + \sigma_2^2)$. Likewise, $Y = Y_1 + Y_2 \sim \ncalN(0, \sigma_1^2 + \sigma_2^2)$. We can now represent $F_1$ and $F_2$ in the abscissa and ordinate axes as $[X_1, Y_1]$ and $[X_2, Y_2]$, respectively. Hence, $F = \sqrt{X ^ 2 + Y ^ 2}$ will have a Rayleigh distribution with parameter $\sigma = \sqrt{\sigma_1^2 + \sigma_2^2}$. This completes the proof.
\hfill 
\IEEEQED
\vspace{-0.3cm}
\subsection{Proof of Proposition \ref{prop:JointSZ}} \label{app:Proposition1}
Similar to the proof of Lemma \ref{lem:RW_Psi_Uniform}, we define $2(n-1)$ auxiliary random variables $X_i = R_i$ and $\Xi_i = \Theta_i$, $1 \leq i \leq n-1$ and find the joint pdf of these $2n$ random variables. Solving this system of $2n$ equations, we can write $R_i$ and $\Theta_i$, $1 \leq i \leq n$ in terms of $S_n$, $Z_n$, $X_i$, and $\Xi_i$, $1 \leq i \leq n-1$. Note that $(2n-1)$ of these equations have trivial solutions, i.e., $R_i = X_i$ and $\Theta_i = \Xi_i$ for $1 \leq i \leq n-1$ and $R_n \!=\! S_n \!-\! \sum_{i=1}^{n-1} X_i$. For $\Theta_n$, we write the governing equation as
\begin{align}\label{GovernTheta_n}
\sum_{i=1}^{n-1} X_i \cos(\Theta_n - \Xi_i) = \frac{Z_n^2 - R_n^2 - \sum_{i, j=1}^{n-1}X_i X_j \cos(\Xi_i - \Xi_j)}{2R_n}.
\end{align}
Note that the exact expression for $\Theta_n$ will not be required for the derivation of our joint pdf of interest. We then derive the determinant of the Jacobian matrix as
\begin{align}\label{JacobianSZ}
|\ncalJ| = \left| \frac{\partial \Theta_n}{\partial Z_n} \right| = \frac{Z_n}{R_n \left| \sum_{i=1}^{n-1} X_i \sin(\Theta_n - \Xi_i) \right|}.
\end{align}
Solving \eqref{GovernTheta_n} and \eqref{JacobianSZ}, we can eliminate $R_n$ and $\Theta_n$ to get $|\ncalJ|$ only in terms of $S_n$, $Z_n$, $X_i$, and $\Xi_i$, $1 \leq i \leq n-1$. Hence, we can write the joint distribution of these $2n$ random variables as
\begin{align*}
f_{S_n, Z_n, {\boldsymbol X}, {\boldsymbol \Xi}}(s, z, {\boldsymbol x}, {\boldsymbol \xi}) = \frac{4z}{(2\pi)^n} \frac{f_R(s - J_{n}) \prod_{i=1}^{n-1} f_R(x_i)}{\sqrt{\left( z^2 - ( s - J_{n} - K_{n} )^2 \right)\left( ( s - J_{n} + K_{n} )^2 - z^2 \right)}},
\end{align*}
where $J_{n} = \sum_{i=1}^{n-1} x_i$, $K_{n} = \sqrt{\sum_{i, j=1}^{n-1}x_i x_j \cos(\xi_i - \xi_j)}$. Integrating this pdf $2(n-1)$ times with respect to $x_i$ and $\xi_i$, $1 \leq i \leq n-1$, we end up with \eqref{JointSZ} and the proof is complete.
\hfill 
\IEEEQED
\vspace{-0.3cm}
\subsection{Proof of Proposition \ref{prop:RWDistributionLt}} \label{app:Proposition2}
Considering the RW mobility model at time $t$, a DBS is either in its first flight, or in its second flight, etc. Hence, initializing $S_0 = 0$, the event $\Omega = \bigcup_{n=1}^\infty \left(S_{n-1} \leq vt < S_n\right)$ has unit probability. Thus, we can write the cdf of $L(t)$ as
\begin{align}\label{IntermediateProofProp2}
F_L(l; t) &= \nbbP[L(t) \leq l] = \nbbP[L(t) \leq l, \Omega] = \sum_{n=1}^\infty \nbbP\left[L(t) \leq l, S_{n-1} \leq vt < S_n\right]\nonumber\\
&\hspace{-1.12cm}\overset{(a)}{=} \sum_{n=1}^\infty \nbbP\left[Z_{n-1}^2 + (vt - S_{n-1})^2 - 2Z_{n-1}(vt - S_{n-1})\cos(\Phi_n) \leq l^2, S_{n-1} \leq vt < S_n\right]\nonumber\\
&\hspace{-1.12cm}\overset{(b)}{=} \sum_{n=1}^\infty \int_0^{vt}\int_0^s \scalebox{0.87}{$ f_{S_{n - 1}, Z_{n - 1}}(s, z)(1 - F_R(vt - s)) \nbbP\left[z^2 + (vt - s)^2 - 2z(vt - s)\cos(\Phi_n) \leq l^2\right]\, {\rm d}z \, {\rm d}s,$}
\end{align}
where in $(a)$ we used the cosine law and in $(b)$ we conditioned the probability on knowing $S_{n-1}$ and $Z_{n-1}$. Note that $\Phi_n = \pi - \Theta_n + \Psi_{n-1}$ is the angle between the direction of $Z_{n-1}$ and the direction of the $n$-th flight. We now rewrite the probability in \eqref{IntermediateProofProp2} as $\nbbP\left[\cos(\Phi_n) \geq x\right]$, where $x = \frac{z^2 + (vt - s)^2 - l^2}{2z(vt - s)}$. In order for this probability to be non-zero, we have two cases: (i) $x \leq -1$ which gives $z + (vt - s) \leq l$, and (ii) $-1 < x \leq 1$ which gives $|z - (vt - s)| \leq l < z + (vt - s)$, which is the triangle inequality. Note also that it is clear from our setup that $z \leq s \leq vt$. When $l \geq vt$, only the first case will hold and we can write the cdf as
\begin{align*}
F_L(l; t) &= \sum_{n=1}^\infty \int_0^{vt}\int_0^s f_{S_{n - 1}, Z_{n - 1}}(s, z)(1 - F_R(vt - s))\, {\rm d}z \, {\rm d}s\\
&= \sum_{n=1}^\infty \int_0^{vt} f_{S_{n - 1}}(s)(1 - F_R(vt - s))\, {\rm d}s = \sum_{n=1}^\infty F_{S_{n - 1}}(vt) - F_{S_{n}}(vt) = 1,
\end{align*}
where in the last equality we used the convolution integral that arises in the derivation of the cdf of $S_n = S_{n-1} + R_n$. Hence, $F_L(l; t) = 1$ for $l \geq vt$, as expected. On the other hand, both cases can occur when $l < vt$. For the first case, we have $0 \leq z \leq l - (vt - s)$ and $vt - l \leq s \leq vt$. Similarly, for the second case we have $|l - (vt - s)| \leq z \leq \min\{s,\, l + (vt - s)\}$ and $\frac{vt - l}{2} \leq s \leq vt$ based on the triangle inequality. Now, according to Lemma \ref{lem:RW_Psi_Uniform}, $\Psi_{n-1}$ is uniformly distributed in $[0, 2\pi)$, and since $\Theta_n \sim [0, 2\pi)$ is independent of $\Psi_{n-1}$, the random variable $\Phi_n$ will have a symmetric triangular distribution. However, since the range of values of $\Phi_n$ is between $0$ and $2\pi$, we have $\Phi_n \sim [0, 2\pi)$. Hence, the cdf of $L(t)$ for $l < vt$ can be written as in \eqref{RWcdfLt}. Note that when $n=1$, we have $L(t) = vt$ and the cdf becomes $\nbbP[vt \leq l, vt < R_1] = (1 - F_R(vt)){\bf 1}(l - vt)$. Differentiating the derived cdf with respect to $l$ using the Leibniz integral rule for two-dimensional integrals, we end up with the pdf of $L(t)$ as written in \eqref{RWpdfLt}.
\hfill 
\IEEEQED
\vspace{-0.3cm}
\subsection{Proof of Proposition \ref{prop:RWPDistributionLt}} \label{app:Proposition3}
Define $M_n = S_{n} + vW_{n-1}$ and $Y_n = S_{n} + vW_{n}$. Considering the RWP mobility model at time $t$, a DBS is either in one of its flight states, or in one of its waiting time states. Hence, initializing $S_0 = 0$ and $W_{-1} = 0$, the event $\Omega' = \bigcup_{n=1}^\infty \left(V_{n-1} \cup F_n\right)$ has unit probability, where $V_n = \{ M_n \leq vt < Y_n\}$ and $F_n = \{Y_{n-1} \leq vt < M_n\}$ are the $n$-th waiting and flight periods, respectively. Thus, we can write the cdf of $L(t)$ as follows.
\begin{align*}
F_L(l; t) =\, &\nbbP[L(t) \leq l] = \nbbP[L(t) \leq l, \Omega'] = \sum_{n=1}^\infty \nbbP\left[L(t) \leq l, V_{n-1}\right] + \sum_{n=1}^\infty \nbbP\left[L(t) \leq l, F_n\right]\nonumber\\
\overset{(a)}{=} \, &\sum_{n=1}^\infty \scalebox{0.92}{$\nbbP\left[Z_{n-1} \leq l, V_{n-1}\right]$} + \sum_{n=1}^\infty \scalebox{0.92}{$\nbbP\left[Z_{n-1}^2 + (vt - Y_{n-1})^2 - 2Z_{n-1}(vt - Y_{n-1})\cos(\Phi_n) \leq l^2, F_n\right]$}\nonumber\\
= \, &\sum_{n=1}^\infty \int_0^{vt}\int_0^{\min\{s, l\}}  f_{S_{n - 1}, Z_{n - 1}}(s, z) \nbbP\left[ s + vW_{n-2} \leq vt < s + vW_{n-2} + vT_{n-1} \right]\, {\rm d}z \, {\rm d}s \, + \nonumber\\
\, &\sum_{n=1}^\infty \int_0^{vt} \int_0^y \scalebox{0.93}{$ f_{Y_{n - 1}, Z_{n - 1}}(y, z) (1 \!-\! F_R(vt \!-\! y)) \nbbP\left[z^2 \!+\! (vt \!-\! y)^2 \!-\! 2z(vt \!-\! y)\cos(\Phi_n) \!\leq\! l^2\right] \, {\rm d}z \, {\rm d}y $} \nonumber\\
\overset{(b)}{=} \, &\sum_{n=1}^\infty \int_0^{vt}\int_0^{\min\{s, l\}}  f_{S_{n - 1}, Z_{n - 1}}(s, z)  \left( F_{W_{n-2}}(t - \frac{s}{v}) - F_{W_{n-1}}(t - \frac{s}{v}) \right)\, {\rm d}z \, {\rm d}s \, + \nonumber\\
\, &\sum_{n=1}^\infty \int_0^{vt} \int_0^y \int_0^{\frac{y}{v}} f_{W_{n - 1}}(w) f_{S_{n - 1}, Z_{n - 1}}(y - vw, z) (1 - F_R(vt - y))\, \times\nonumber\\
&\hspace{3cm} \nbbP\left[z^2 + (vt - y)^2 - 2z(vt - y)\cos(\Phi_n) \leq l^2\right] \, {\rm d}w \, {\rm d}z \, {\rm d}y,
\end{align*}
where in $(a)$ we used the fact that $L(t) = Z_{n-1}$ when a DBS is in its $(n-1)$-th waiting state and $L(t)$ follows the cosine rule when a DBS is in its $n$-th flight state. In the first summation of $(b)$, we first conditioned the probability in the integrand on knowing $T_{n-1}$ and then used the identity $W_{n-1} = W_{n-2} + T_{n-1}$ to simplify the result. In the second summation of $(b)$, we used the definition of $Y_{n-1}$ to write the joint pdf of $Y_{n-1}$ and $Z_{n-1}$ in terms of the joint pdf of $S_{n-1}$ and $Z_{n-1}$. Now, with the same reasoning as in the proof of Proposition \ref{prop:RWDistributionLt}, we get the cdf and pdf of $L(t)$ when $l \leq vt$ as in \eqref{RWPcdfLt} and \eqref{RWPpdfLt}, respectively. Note that for $n=1$, we have
\begin{align*}
F_L(l; t | n = 1) &= \nbbP[vt \leq vW_0] + \nbbP[vt \!-\! vW_0 \leq l, vW_0 \leq vt < R_1 \!+\! vW_0]= \nbbP\!\left[\scalebox{0.92}{$W_0 \geq t \!-\! \frac{\min\{R_1, l\}}{v}$}\right]  \\
&= \int_0^\infty \nbbP[\min\{R_1, l\} \geq vt - vw] f_T(w)\, {\rm d}w = \int_{t - \frac{l}{v}}^\infty \left(1 - F_R(vt - vw)\right)f_T(w)\, {\rm d}w.
\end{align*}
When $l > vt$, we can write the cdf as
\begin{align*}
F_L(l; t) =\, &\sum_{n=1}^\infty \! \int_0^{vt} \!\! \scalebox{0.92}{$ f_{S_{n - 1}}(s)\left( F_{W_{n-2}}(t - \frac{s}{v}) - F_{W_{n-1}}(t - \frac{s}{v}) \right) \, {\rm d}s$} + \sum_{n=1}^\infty \! \int_0^{vt}\!\!  \scalebox{0.92}{$f_{Y_{n - 1}}(y)(1 - F_R(vt - y))\, {\rm d}y$}\\
\overset{(a)}{=}\, &\sum_{n=1}^\infty \!\int_0^{vt} \!\!\! \scalebox{0.9}{$ f_{S_{n - 1}}(s) F_{W_{n-2}}(t - \frac{s}{v})\, {\rm d}s $} -\!\! \int_0^{vt} \!\!\! \scalebox{0.9}{$ f_{Y_{n - 1}}(y) F_R(vt - y)\, {\rm d}y $} \overset{(b)}{=} \sum_{n=1}^\infty \scalebox{0.9}{$ F_{M_{n - 1}}(vt) \!-\! F_{M_{n}}(vt) =\! 1$}
\end{align*}
where in $(a)$ we used the definition of $Y_{n-1}$ to get $F_{Y_{n-1}}(vt) = \int_0^{vt} f_{S_{n - 1}}(s) F_{W_{n-1}}(t - \frac{s}{v})\, {\rm d}s$ and in $(b)$ we used the definition of $M_{n-1}$ and the identity $M_n = Y_{n-1} + R_n$ to get the result.
\hfill 
\IEEEQED

\vspace{-0.4cm}
\bibliographystyle{IEEEtran}
\bibliography{J1}

\end{document}